%% file: effphafielderr_SIAP.tex
\documentclass[final]{siamart0216}

\usepackage{lipsum}
\usepackage{amsfonts}
\usepackage{graphicx}
\usepackage{epstopdf}
\usepackage{algorithmic}
\newcommand{\reff}[1]{(\ref{#1})}

\newcommand{\cg}[1]{\mathcal{#1}}

\newcommand{\av}[1]{\left|#1\right|}

\newcommand{\Ll}[1]{\left\|#1\right\|}
\newcommand{\N}[2]{\left\|#1\right\|_{#2}}

\newcommand{\brkts}[1]{\left(#1\right)}
\newcommand{\ebrkts}[1]{\left[#1\right]}

\newcommand{\brcs}[1]{\left\{#1\right\}}

\newcommand{\pd}[2]{\frac{\partial #1}{\partial #2}}

\newcommand{\bsplitl}[2]{
\begin{equation}
\begin{split}
#1
\end{split}
\label{#2}
\end{equation}}
\newcommand{\bsplit}[1]{
\begin{equation*}
\begin{split}
#1
\end{split}
\end{equation*}}

\newcommand{\cblu}[1]{{\color{black} #1}}

\ifpdf
  \DeclareGraphicsExtensions{.eps,.pdf,.png,.jpg}
\else
  \DeclareGraphicsExtensions{.eps}
\fi

\newtheorem{thm}{Theorem}
\newtheorem{lem}{Lemma}
\newtheorem{rem}{Remark}
\newtheorem{defn}{Definition}

\newcommand{\TheTitle}{Rate of Convergence of General Phase Field Equations towards their Homogenized Limit} 
\newcommand{\TheAuthors}{M. Schmuck \and S. Kalliadasis}

\headers{Error estimates for upscaled phase field equations}{\TheAuthors}

\title{{\TheTitle}\thanks{This work was funded by Engineering and Physical Sciences
Research Council of the UK through Grants Nos. EPSRC Grant Nos. EP/H034587/1,
EP/L027186/1, EP/L025159/1, EP/L020564/1, EP/K008595/1, and 
EP/P011713/1 and from the European Research Council via Advanced Grant No. 247031.}}

\author{
	M. Schmuck \thanks{
	Maxwell Institute for Mathematical Sciences and 
	School of Mathematical and Computer Sciences,
	Heriot-Watt University,
	EH14 4AS, Edinburgh, UK, 
	(\email{M.Schmuck@hw.ac.uk}, 
	\url{http://www.macs.hw.ac.uk/\~ms713}).}
	\and
	S. Kalliadasis \thanks{
	Department of Chemical Engineering,
	Imperial College London,
	South Kensington Campus,
	SW7 2AZ London, UK, 
	(\email{s.kalliadasis@imperial.ac.uk}).}
}

\usepackage{amsopn}

\begin{document}

\maketitle

\begin{abstract}
Over the last few decades, phase-field equations have found increasing
applicability in a wide range of mathematical-scientific fields (e.g.
geometric PDEs and mean curvature flow, materials science for the study of
phase transitions) but also engineering ones (e.g. as a computational tool in
chemical engineering for interfacial flow studies). Here, we focus on
phase-field equations in strongly heterogeneous materials with perforations
such as porous media. To the best of our knowledge, we provide the first derivation of error estimates for
fourth order, homogenized, and nonlinear evolution equations. Our fourth order 
problem induces a slightly lower convergence rate, i.e., 
$\epsilon^{1/4}$, where $\epsilon$ denotes the material's specific
heterogeneity, than established for second-order elliptic problems (e.g. \cite{Zhikov2006})
for the error between the effective macroscopic solution of the (new)
upscaled formulation and the solution of the microscopic phase field problem.
We hope that our study will motivate new modelling, analytic, and computational 
perspectives for interfacial transport and phase transformations in strongly
heterogeneous environments.
\end{abstract}



\section{Introduction}\label{sec:Intr}
\input{realIntro9.tex}

In Section \ref{sec:2Fo}, we present basic notations and mathematical assumptions. The main results are summarized in Section \ref{sec:MaRe} and subsequently justified
in Sections \ref{sec:FoUp} and \ref{sec:ErEs}. Conclusions and suggestions for further work are given in
Section~\ref{sec:Concl}.

\section{Mathematical preliminaries and notation}\label{sec:2Fo}
\input{reform7.tex}

\section{Main results}\label{sec:MaRe}
\input{equilibrium4.tex}
\input{mainresults7.tex}

\section{Formal derivation of upscaled equations}\label{sec:FoUp}
\input{upscaling4.tex}

\section{Proof of Theorem \ref{thm:ErEs}}\label{sec:ErEs}
\input{error.tex}

\section{Conclusions}\label{sec:Concl}
\input{conclusion8.tex}


\section*{Acknowledgments}
We would like to thank the Referees for their helpful comments and 
careful reading of the manuscript.

\bibliographystyle{siamplain}
\bibliography{effWetting7}
\end{document}

%% file: realIntro9.tex
We consider the well-accepted Cahn-Hilliard/diffuse-interface formulation
\cite{Cahn1958,vanDerWaals1892} for studying the evolution of interfaces
between different phases. Its broad applicability together with increasing
computational power have enabled its use to new and increasingly complex
scientific and engineering problems such as the computation of transport
equations in porous media \cite{Sahimi2011} which represents a numerically
very demanding, high-dimensional multiscale problem \cite{Jenny2003}. The
purpose of the present work is to rigorously and systematically provide a
reliable effective macroscopic description of how multiple phases invade
strongly heterogeneous media, such as porous materials for instance.

The cornerstone of phase-field models is the abstract energy density
\bsplitl{
e(\phi)
    := \frac{1}{\lambda}F(\phi) 
    +\frac{\lambda}{2}\av{\nabla\phi}^2\,,
}{FrEn}
where $\phi:=\frac{c_\beta}{c_\alpha+c_\beta}$ is a reduced order parameter
representing the fraction of species of type $\beta$ in a binary solution containing species
$\alpha$ and $\beta$ with number densities $c_\alpha$ and $c_\beta$, respectively. The gradient term
$\cblu{\lambda/2}\av{\nabla\phi}^2$ penalizes the interfacial area between
these phases, and $F$ is defined as the general
(Helmholtz) free energy density
%
$F(\phi)
    := U-TS$\,,
%
where $U$ is the internal energy, $T$ is the temperature and $S$ is the
entropy. \cblu{The parameter $\lambda>0$ is proportional to the 
interfacial width and leads to the appearance of smooth interface.}

Important examples of this formulation include the regular solution theory which
has been applied successfully in a wide spectrum of scientific and
technological contexts such as ionic melts \cite{Hillert1970}, water sorption
in porous solids \cite{Bazant2012}, and micellization in binary surfactant
mixtures \cite{Huang1997}. The key quantity in this theory is the so-called
regular solution energy density (also known as the Flory-Huggins energy
density \cite{Flory1942})
%
$F(\phi)
	:= R(\phi)
	-TS_I(\phi)
	$\,,
%
where $S_I(\phi) := -k_B\ebrkts{
	\phi{\rm ln}\,\phi-(1-\phi){\rm ln}\,(1-\phi) }$ is the entropy of mixing
for ideal solutions and the regular solution term
$R(\phi):=z\omega\phi(1-\phi)$ accounts for the interaction energy between
different species. The variable $z$ is the coordination number defining the
number of bonds of $\beta$ with neighbouring species.
$\omega:=\epsilon_{\alpha\alpha}+\epsilon_{\beta\beta}-2\epsilon_{\alpha\beta}$
is the interaction energy parameter accounting for the minima
$\epsilon_{\alpha\alpha}$, $\epsilon_{\beta\beta}$, and
$\epsilon_{\alpha\beta}$ of interaction potentials which define attractive
and repulsive forces between the species $\alpha$ and $\beta$.

Wetting phenomena, often studied using classical sharp-interface
approximations, e.g.~\cite{Nikos2009,Nikos2010,Nikos2011,Raj2011}, also enjoy
a wide-spread use of phase-field
modeling~\cite{Pomeau2001,Yue2010,Wylock2012,David2013,David2013b} even in
the presence of complexities such as an electric field (so called
electrowetting, e.g.~\cite{Eck2009,Nochetto2013}). The reason for this is
that classical sharp-interface models consider the fluid-fluid interface to
be a sharp surface of zero thickness where quantities such as the fluid
density are, in general, discontinuous, which leads to singularity formation
for interfacial problems with topological transitions, e.g. the notorious
contact line singularity~\cite{Huh1971} often cured with phenomenological
approaches such as slip models. The phase-field/diffuse-interface approach
relaxes the assumption of a sharp interface in line with the physics of the
problem and in agreement with developments and applications in the field of
statistical mechanics of liquids and in molecular simulations, with
quantities varying smoothly but rapidly, and considers the interface to have
a non-zero thickness, thus allowing a ``natural" regularisation for
singularities in interfacial problems with topological transitions.

Other applications include transport in electrochemical systems e.g.
consisting of an electrolyte and an electrode \cite{Guyer2004}, or immiscible
flows \cite{Liu2003,Otto1997} under a polynomial free energy in the form of
the classical double-well potential, i.e., $W(\phi):=\frac{1}{4}(1-\phi^2)^2$
are relevant applications. Phase-field energy functionals are also of
interest in image processing such as inpainting, see
e.g.~\cite{Bertozzi2007}.

Our formal derivation of upscaled phase-field equations is valid for general
free energies but the subsequent rigorous derivation of error estimates is
based on free energies of the following form.

{\bf Polynomial Class (PC):} \emph{Admissible free energy densities $F$ in \reff{FrEn} are polynomials
of order $\cblu{2r}$, i.e.,
\bsplitl{
\cblu{
F(u)
	= \sum_{i=2}^{2r}b_iu^i\,,
	\quad ib_i=a_{i-1}\,,
	\quad 2\leq i\leq 2r
	}
	\,,
}{PolyDef}
with $f(u)=F' (u)$ vanishing at $u=0$, that is,
\bsplitl{
\cblu{
f(u)
	= \sum_{i=1}^{2r-1}a_iu^i\,,
	\quad r\in\mathbb{N}\,,
	\quad r\geq 2
	}
	\,,
}{PNfrEn}
where the leading coefficient of both $F$ and $f$ is positive, i.e.,
$a_{2r-1}=2rb_{2r}>0$.
}

\medskip

Temam \cite{Temam1997} established well-posedness of the Cahn-Hilliard equation for free
energies of class (PC). In computations, one often replaces the regular
solution energy density, composed of $R$ and $S_I$ defined above, by
the polynomial double-well potential $W(\phi)$.

In difference to \cite{Schmuck2012b}, we provide here an upscaling strategy
that is valid for general homogeneous free energy densities by making use of
a Taylor expansion of the free energy density at the effective upscaled
solution. This serves also as a general methodology for the homogenization of
nonlinear problems. Moreover, to the best of our knowledge, we present here
for the first time, error estimates between the solution of the microscopic
phase-field equations solved in a periodic porous medium and the solution of
the correspondingly homogenized/upscaled equations by Theorem \ref{thm:ErEs} below.
In the remaining part of this section, we introduce the basic equations
describing interfacial dynamics in a homogeneous environment and subsequently in a periodic porous medium.

{\bf (a) Homogeneous domains $\Omega$.} In the Ginzburg-Landau/Cahn-Hilliard
formulation, the total energy is defined by $E(\phi):=\int_\Omega
e(\phi)\,d{\bf x}$ with density \reff{FrEn} on a bounded domain
$\Omega\subset\mathbb{R}^d$ with smooth boundary $\partial\Omega$ and $1\leq
d\leq 3$ denotes the spatial dimension. It is well accepted that
thermodynamic equilibrium can be achieved by minimizing the energy $E$,
frequently supplemented by a wetting boundary contribution $\int_{\partial\Omega}g({\bf
x})\,d{\bf x}$ for $g({\bf x})\in H^{3/2}(\partial\Omega)$.
The wetting property of pore walls can be characterised by
$$
g({\bf x})=-\frac{\gamma}{C_h}a({\bf x})\,,
$$
where $C_h$ is the Cahn number \cblu{$\frac{\lambda}{L}$, $L$ the macroscopic length scale, and 
$\gamma=\frac{2\sqrt{2}\phi_e}{3\sigma_{lg}}$}, $\sigma_{lg}$ is the liquid-gas
surface tension, and $\phi_e$ is the local equilibrium limiting value of $F$,
see \cite{Schmuck2012}. For simplicity, we set subsequently $g=0$ and hence
assume walls with neutral wetting characteristics, i.e., walls inducing a
contact angle of 90 degrees. A widely used minimization over time forms the
$H^{-1}$-gradient flow with respect to $E(\phi)$, i.e., \bsplitl{
\textrm{(Homogeneous case)}\,\,\,
\begin{cases}
\pd{}{t}\phi
    = {\rm div}\brkts{
    \hat{\rm M}\nabla\brkts{
       \frac{1}{\lambda} f(\phi) 
        -\lambda\Delta\phi
        }
    }
    & \quad\textrm{in }\Omega_T\,,
\\
\nabla_n\phi:= {\bf n}\cdot\nabla\phi
    = g({\bf x})
    & \quad\textrm{on }\partial\Omega_T 
    \,,
\\
\nabla_n\Delta\phi
    = 0
    & \quad\textrm{on }\partial\Omega_T
    \,,
\end{cases}
}{PhMo}
where
$\Omega_T:=\Omega\times]0,T[$, $\partial\Omega_T:=\partial\Omega\times]0,T[$, $\phi$ satisfies the initial condition $\phi({\bf x},0)
    = \psi({\bf x})$, and $\hat{\rm M}=\brcs{{\rm m}_{ij}}_{1\leq i,j\leq d}$ denotes \cblu{a symmetric and positive definite} mobility tensor. \cblu{Throughout the 
article we write $]a,b[$ for open intervals with $a,b\in\mathbb{R}$ and $a<b$.} 
The gradient flow \reff{PhMo} is weighted by the mobility tensor $\hat{\rm
M}$, and is referred to as the Cahn-Hilliard equation. This equation is a
model prototype for interfacial dynamics, e.g. \cite{Fife1991}, and phase
transformation, e.g. \cite{Cahn1958}, under homogeneous Neumann boundary
conditions, i.e., $g=0$, and free energy densities $F$.

We recall that the integrated energy density \reff{FrEn} dissipates along
solutions of \reff{PhMo}, that means,
$E(\phi(\cdot,t))
    \leq
    E(\phi(\cdot,0))
    =:E_0
$. This follows
immediately after differentiating $E(\phi)$ with respect to time and using
\reff{PhMo} for $g=0$.

There is also an interesting connection between the Cahn-Hilliard/phase-field
equation and the free-boundary value problem known as the Mullins-Sekerka
problem \cite{Mullins1964} or the two-phase Hele-Shaw problem
\cite{HeleShaw1898}.
The Hele-Shaw problem plays a crucial role for deriving
more regular solutions of the Cahn-Hilliard equation \reff{PhMo}, see \cite{Alikakos1994}.
Inspired by the formal
derivation by Pego \cite{Pego1989}, it was rigorously verified later on in
\cite{Alikakos1994,Soner1995} that the chemical potential
\bsplitl{
\mu(\phi)
	:=-\lambda\Delta\phi+\frac{1}{\lambda}f(\phi)
	\,,
}{chemPot}
satisfies 
\cblu{for an evolving interfacial front $\Gamma_t$ with initial condition 
$\Gamma_{00}$} in the 
limit $\lambda\to 0$ for $t\in [0,T]$ the following
\bsplitl{
\hspace{-0.35cm}
\textrm{Hele-Shaw/Mullins-Sekerka problem:}\,\,
\begin{cases}
\quad
\Delta \mu
	= 0
	&\quad\textrm{in }\Omega\setminus\Gamma_t\,,
\\\quad
{\bf n}\cdot\nabla\mu
	= 0
	&\quad\textrm{on }\partial\Omega\,,
\\\quad
\mu
	= \sigma\kappa
	&\quad\textrm{on }\Gamma_t\,,
\\\quad
v = \frac{1}{2}\ebrkts{{\bf n}\cdot\nabla\mu}_{\Gamma_t}
	&\quad\textrm{on }\Gamma_t\,,
\\\quad
\Gamma_0
	= \Gamma_{00}
	&\quad\textrm{if }t=0\,,
\end{cases}
}{HSP}
where $\sigma=\int_{-1}^{1}\brkts{\frac{1}{2}\int_0^sf(r)\,dr}^{1/2}\,ds$ is the interfacial tension, $\kappa$ the mean curvature,
$v$ the normal velocity of the interface $\Gamma_t$, ${\bf n}$ the unit outward normal to either $\partial\Omega$ or $\Gamma_t$, and
$\ebrkts{{\bf n}\cdot\nabla\mu}_{\Gamma_t}:={\bf n}\cdot\nabla\mu^+-{\bf n}\cdot\nabla\mu^-$ where $\mu^+:=\mu\,\bigr|_{\Omega^+_t}$ and $\mu^-:=\mu\,\bigl|_{\Omega^-_t}$ and $\Omega^+_t$ and $\Omega^-_t$ denote the exterior and interior of
$\Gamma_t$ in $\Omega$. Herewith, we also have $\phi\to \pm 1$ in $\Omega^\pm_t$ for all $t\in [0,T]$ as $\lambda\to 0$. Finally, the
derivation of convergence rates (Theorem~\ref{thm:ErEs} below) requires higher regularity of solutions
of the Cahn-Hilliard equation (Assumption C below)
than available in \cite{Alikakos1994,FengX2004}, which require the existence of global in time solutions of the sharp interface limit \reff{HSP}.

\begin{figure}
\centering
\includegraphics[width=9cm]{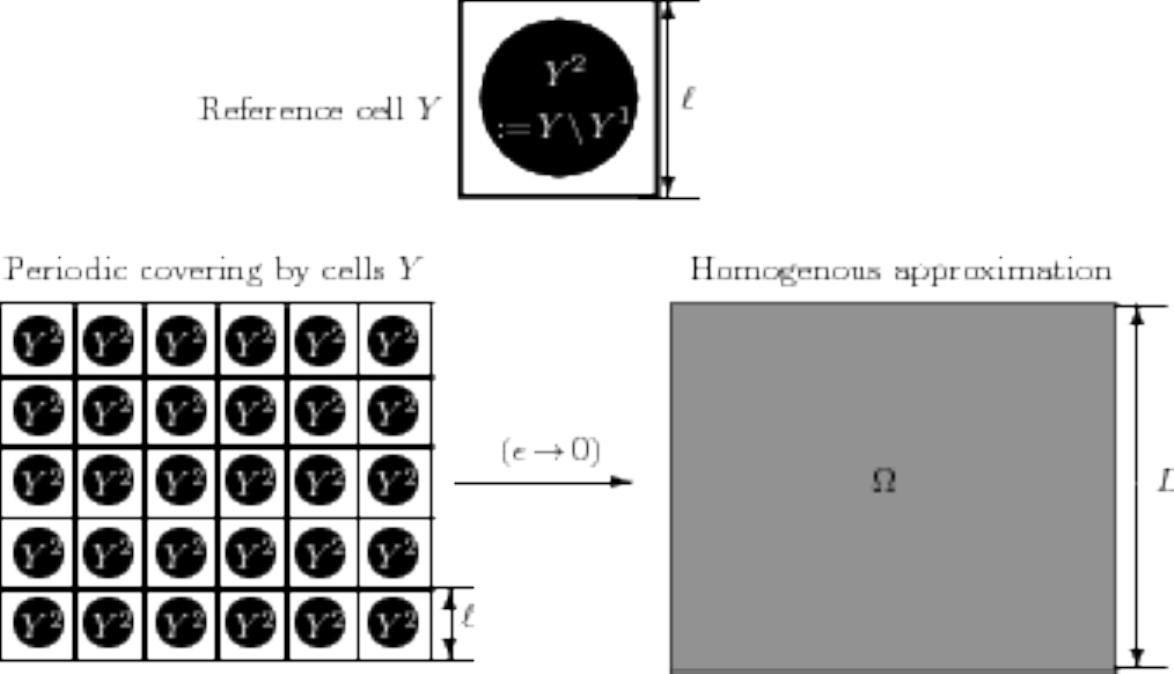}
\caption{{\bf Left:} Strongly heterogeneous/perforated material as a periodic covering of reference cells $Y:=[0,\ell]^d$.
{\bf Top, middle:} Definition of the reference cell $Y=Y^1\cup Y^2$ with $\ell=1$.
{\bf Right:} The ``homogenization limit'' $\epsilon:=\frac{\ell}{L}\to 0$ scales the perforated domain such that perforations
become invisible on the macroscale.}
\label{fig:MicMac}
\end{figure}

{\bf (b) Heterogeneous/perforated domains $\Omega^\epsilon$.} Our main study
concentrates on \reff{FrEn} in perforated domains
$\Omega^\epsilon\subset\mathbb{R}^d$ instead of a homogeneous domain
$\Omega\subset\mathbb{R}^d$. The dimensionless variable $\epsilon>0$ defines
the heterogeneity $\epsilon =\frac{\ell}{L}$ where $\ell$ represents the
characteristic pore size and $L$ is the macroscopic length of the porous
medium, see Figure \ref{fig:MicMac}. Hence, the porous medium is defined by a
reference pore/cell $Y:=
[0,\ell_1]\times[0,\ell_2]\times\dots\times[0,\ell_d]$. For simplicity, we
set $\ell_1=\ell_2=\dots=\ell_d=1$.
The pore and the solid phase
of the medium are denoted by $\Omega^\epsilon$ and $B^\epsilon$,
respectively. These sets are defined by,
\bsplitl{
\Omega^\epsilon
    & := \bigcup_{{\bf z}\in\mathbb{Z}^d}\epsilon\brkts{Y^1+{\bf z}}\cap\Omega\,,
\qquad
B^\epsilon
    := \bigcup_{{\bf z}\in\mathbb{Z}^d}\epsilon\brkts{Y^2+{\bf z}}\cap\Omega
    =\Omega\setminus\Omega^\epsilon\,,
}{Oe2}
where the subsets $Y^1,\,Y^2\subset Y$ are such that $\Omega^\epsilon$
is a connected set. More precisely, $Y^1$ stands for the pore phase (e.g. liquid or gas phase in wetting problems),
see Figure \ref{fig:MicMac}. Additionally, we define the macroscopic pore walls by
$I_\Omega^\epsilon:=\partial\Omega^\epsilon\cap\partial B^\epsilon$ and the microscopic pore
walls by $I_Y:=\partial Y^1\cap\partial Y^2$.
Herewith, we can reformulate \reff{PhMo} for $g=0$ by the
following microscopic porous media problem
\bsplitl{
\textrm{(Micro porous case)}\,\,\,
\begin{cases}
\quad \partial_t\phi_\epsilon
    = {\rm div}\brkts{
    \hat{\rm M}\nabla \brkts{
            -\lambda\Delta \phi_\epsilon
            +\frac{1}{\lambda}f(\phi_\epsilon)
        }
    }
    & \quad\textrm{in }\Omega^\epsilon_T\,,
\\\quad
\nabla_n\phi_\epsilon:= {\bf n}\cdot\nabla\phi_\epsilon
    = 0
    & \quad\textrm{on }\partial\Omega^\epsilon_T
    \,,
\\\quad
\nabla_n\Delta\phi_\epsilon
    = 0
    & \quad\textrm{on }\partial\Omega^\epsilon_T
    \,,
\\\quad
\phi_\epsilon({\bf x},0)
    = \psi({\bf x})
    & \quad\textrm{on }\Omega^\epsilon\,.
\end{cases}
}{PeMoPr}
Our main objective is the derivation of error estimates for the
difference between the upscaled/homogenized solution
$\phi_0$ of \reff{pmWrThm} and the microscopic solution $\phi_\epsilon$ of
\reff{PeMoPr} in order to have a \emph{qualitative and quantitative measure
for the validity} of the homogenized phase field formulation \reff{pmWrThm}
(Theorem \ref{thm:ErEs}) obtained by passing to the limit $\epsilon\to 0$ in
\reff{PeMoPr}. This result will also provide a rigorous basis for the formal
upscaling in \cite{Schmuck2014}. The homogenized equation stated in
Theorem~\ref{thm:ErEs} below allows for new analytical considerations such as a sharp interface study of the novel upscaled equation or establishing more regular solutions of Cahn-Hilliard/phase field equations as
well as for new avenues in modelling. It ultimately leads to convenient, low-dimensional computational schemes which
can be solved by well-known numerical methods developed for homogeneous
domains.

%% file: reform7.tex
We recall the splitting formulation of the Cahn-Hilliard
equation from \cite{Schmuck2012b} which builds the basis for our subsequent homogenization analysis. To this end, we set 
$$
H^2_E(\Omega):=\brcs{\phi\in H^k(\Omega)\,\bigl|\,\nabla_n\phi=0\text{ and }\overline{\phi}\cblu{:=\frac{1}{|\Omega|}\int_\Omega\phi\,d{\bf x}}=0,\,\cblu{k\geq 2}}
$$ 
and identify
$\phi
    = (-\Delta)^{-1}w$
in the $H^2_E(\Omega)$-sense, this means, we have for all $\varphi\in H^2_E(\Omega)$ that
$$
\brkts{-\Delta\phi,\varphi}
	= \brkts{-\Delta(-\Delta)^{-1}w,\varphi}
	= \brkts{w,\varphi}
	\,,
$$
\cblu{where $(\cdot,\cdot)$ denotes the standard $L^2$-scalar product.} 
Herewith we can rewrite (\ref{PeMoPr}) (for simplicity stated here for $\Omega\subset\mathbb{R}^d$ instead of $\Omega^\epsilon$) for all $\varphi\in H^2_E(\Omega)$ as
\bsplit{
\brkts{\partial_t(-\Delta)^{-1}w,\varphi}
	-\brkts{\lambda{\rm div}\brkts{
    \hat{\rm M}\nabla
        w
    },\varphi}
	& = \brkts{{\rm div}\brkts{
    \frac{\hat{\rm M}}{\lambda}\nabla
        f(\phi) 
    },\varphi}
    \,,
\\
\brkts{\nabla\phi,\nabla\varphi}
	& = \brkts{w,\varphi}
	\,,
}
which reads in the classical sense
\bsplitl{
\textrm{\bf (Splitting)}\quad
\begin{cases}
\quad \partial_t(-\Delta)^{-1}w
    -\lambda{\rm div}\brkts{
    \hat{\rm M}\nabla
        w
    }
        = {\rm div}\brkts{
    \frac{\hat{\rm M}}{\lambda}\nabla
        f(\phi) 
    }
    & \textrm{in }\Omega_T\,,
\\\quad
\nabla_n w
    = -\nabla_n\Delta\phi
    = 0
    &\textrm{on }\partial\Omega_T\,,
\\\quad
-\Delta \phi
    = w
    &\textrm{in }\Omega_T\,,
\\\quad
\nabla_n \phi
    = 0
    &\textrm{on }\partial\Omega_T\,,
\\\quad
\phi({\bf x},0)
    = \psi({\bf x})
    & \textrm{in }\Omega\,.
\end{cases}
}{DePhMo0}
In \cite{Novick-Cohen1990}, the existence of a local solution $\phi \in
H^2_E(\Omega)$ to equation (\ref{PeMoPr}) has been verified for $f\in
C^2_{Lip}(\mathbb{R})$ and hence also to (\ref{DePhMo0}). Furthermore,
Novick-Cohen \cite{Novick-Cohen1990} states necessary conditions for global
existence while a proof based on Galerkin approximations and a priori
estimates can be found in \cite[Theorem 4.2, p. 155]{Temam1997}. The
advantage of \reff{DePhMo0} is that it allows to base our upscaling approach
on well-known results from elliptic/parabolic homogenization theory
\cite{Bensoussans1978,Cioranescu1999,Hornung1997,Mei2010,Pavliotis2008,Zhikov1994}.
The splitting \reff{DePhMo0} slightly differs from the 
\cblu{strategy of substituting the chemical potential, which is often applied for computational purposes, see \cite{Barrett1999}, and which 
seems also more appropriate for other homogenization strategies 
such as periodic unfolding \cite{Cioranescu2006} or 
two-scale convergence \cite{Allaire1992,Nguetseng1989} 
for instance.} 

\medskip

Next, we briefly summarize what, to the best of our knowledge, we believe to be the best available regularity results (Lemma~\ref{lem:ApEs} below) for the Cahn-Hilliard equation \cite{Alikakos1994,FengX2004}. These results depend on
two assumptions:

\medskip

{\bf Assumption A:}
\emph{
\begin{itemize}
\item[{\rm\bf (A1)}] $F\in C^4(\mathbb{R})$ satisfies $F(\pm 1)=0$ and $F>0$ elsewhere.
\item[{\rm\bf (A2)}] $f(u)=F'(u)$ satisfies for some finite $\alpha>2$ and positive constants $k_i>0$, $i=0,\dots,3$,
\bsplitl{
k_0\av{u}^{\alpha-2}-k_1
	\leq f'(u)
	\leq k_2\av{u}^{\alpha-2}+k_3
	\,.
}{f2}
\item[{\rm\bf (A3)}] There exist constants $0<a_1\leq 1$, $a_2>0$, $a_3>0$ and $a_4>0$ such that for $b\in\mathbb{R}$
\bsplitl{
\hspace{-0.5cm}\brkts{
	\cblu{f(a)-f(b)},a-b
	}
	& \geq a_1 \brkts{
		\cblu{f'(a)}(a-b),a-b
	}
	-a_2\av{a-b}^{2+a_3}
	\quad\forall\av{a}\leq 2a_4\,,
\\
aF''(a)
	& \geq 0
	\qquad\forall\av{a}\geq a_4\,.
}{f3}
\end{itemize}
}

It is straightforward to check that the classical double-well potential $F(x)=(x^2-1)^2/4$ satisfies Assumption A.
The following characterization of the initial condition $\psi$ is also required for more regular solutions as derived in
\cite{Alikakos1994,FengX2004} and stated in Lemma \ref{lem:ApEs} below. 
\cblu{We will frequently write $\Ll{u}$ for the $L^2$-norm of a function 
$u$.}

{\bf Assumption B:}
\emph{There exist uniform constants $m_0,\,\sigma_j>0$, $j=1,2,3$ such that
\begin{itemize}
\item[{\rm\bf (B1)}] \qquad $-1 < m_0
		:= \frac{1}{\av{\Omega}}\int_\Omega \psi({\bf x})\,d{\bf x}<1\,,$
\item[{\rm\bf (B2)}] \qquad ${\cal E}_\lambda(\psi)
		:= \frac{\lambda}{2}\Ll{\nabla \psi}^2
		+\frac{1}{\lambda}\N{F(\psi)}{L^1}
		\leq C\lambda^{-2\sigma_1}\,,$
\item[{\rm\bf (B3)}] \qquad $\N{\omega^\lambda}{H^l}
		:=\N{-\lambda\Delta\psi+\frac{1}{\lambda}F(\psi)}{H^l}
		\leq C\lambda^{-\sigma_{2+l}}\,,\quad l=0,1\,,$
\end{itemize}
\cblu{where $|\Omega|$ is the Lebesgue measure of $\Omega|$}. 
}

Herewith, the following regularity result has been derived for
homogeneous domains $\Omega$ in \cite{Alikakos1994,FengX2004}. 


\begin{lem} \label{lem:ApEs} \emph{(Regularity)} Let $f$ and $\psi$ satisfy the \emph{Assumption A} and \emph{B}, respectively.
Moreover, we suppose that the Hele-Shaw/Mullins-Sekerka problem \reff{HSP} has a global in time classical solution. Then, the solution $\phi$ of the
Cahn-Hilliard equation \reff{PhMo} satisfies the estimates
\bsplitl{
\begin{cases}
\quad
\N{\phi}{L^\infty(\Omega_T)}
	\leq C
	\,,
\\\quad
\int_0^\infty \Ll{\nabla\Delta\phi}^2\,dt
	\leq C(\lambda)
	\,,
\\\quad
\N{\Delta^2\phi}{L^\infty([0,\infty[;L^2(\Omega))}
	\leq C\lambda^{-C}\,,
\end{cases}
}{PhiBnd}
for all $\lambda\in]0,\kappa[$ and a family of smooth initial data $\brcs{\psi^\lambda}_{0<\lambda\leq 1}$
where $\kappa$ and $C$ are constants. Estimate \reff{PhiBnd}$_3$ holds for $C>0$ large enough, if
$\lim_{s\to 0^+}\Ll{\nabla \partial_t\phi(s)}
	\leq C\lambda^{-\kappa}$.
\end{lem}

\begin{rem} \emph{(Hele-Shaw)} Existence and uniqueness of classical solutions for the so-called
single phase Hele-Shaw problem in bounded domains in $\mathbb{R}^d$ can be found for instance in
\cite{Escher1997,Meirmanov2002}. \hfill$\diamond$
\end{rem}

We refer the interested reader to Refs. \cite{Alikakos1994,FengX2004} for  a proof.
Since we need slightly stronger regularity results
than stated in Lemma~\ref{lem:ApEs} for the proof of error estimates
(Theorem~\ref{thm:ErEs}), we introduce the following well-accepted 
(see for instance \cite{Cioranescu1999})

\medskip

{\bf Assumption C:} \emph{
For smooth data, i.e.,
$\phi_0({\bf x},0),\phi_\epsilon({\bf x},0)\in C^\infty(\Omega^\epsilon)$,
$f\in C^\infty(\mathbb{R})$, and for $\Omega^\epsilon$ with Lipschitz
boundary $\partial\Omega^\epsilon$ and hence the interface
$I^\epsilon_\Omega$ is Lipschitz too, then the solutions $\phi_0$ of
equation \reff{pmWrThm} and $\phi_\epsilon$ the solution of the microscopic
equation \reff{PeMoPr} satisfy
\bsplitl{
\phi_0,\phi_\epsilon \in \cblu{C^1(0,T;W^{k,\infty}(\Omega^\epsilon))}
	\quad\text{for a }k\geq 4\,.
}{regA}
Moreover, the correctors $\xi_\phi^k$ and $\xi_w^k$, \cblu{which solve the 
cell problems \reff{pmRCTh},} satisfy
\bsplitl{
\xi_\phi^k,\,\xi_w^k\in W^{1,\infty}(Y^1)
	\quad\text{for all}\quad
	1\leq k\leq d
	\,.
}{regXi}
}

Let $T_\epsilon$ denote the extension operator, which extends the solutions
$\phi^\epsilon$ and $w^\epsilon$ of \reff{DePhMo0} defined on the perforated
domain $\Omega^\epsilon$ to the homogeneous domain $\Omega$.
For convenience we denote these extensions by
$\phi^\epsilon$ and $w^\epsilon$ and skip the extension operator $T_\epsilon$
most of the time. The existence of such an operator
$T_\epsilon\,:\,W^{1,p}(\Omega^\epsilon)\to W^{1,p}_{loc}(\Omega)$ for
$\epsilon>0$ was established in \cite{Acerbi1992} and $T_\epsilon$ is characterized by the
following properties:
\bsplitl{
\begin{cases}
\quad
{\bf (T1)}
	& \qquad
	T_\epsilon u =u\quad\textrm{a.e. in }\Omega^\epsilon\,,
\\\quad	
{\bf (T2)}
	& \qquad
	\int_{\Omega(\epsilon k_0)}\av{T_\epsilon u}^p\,d{\bf x}
	\leq k_1\int_{\Omega^\epsilon}\av{u}^p\,d{\bf x}\,,
\\\quad
{\bf (T3)}
	&\qquad
	\int_{\Omega(\epsilon k_0)}\av{D(T_\epsilon u)}^p\,d{\bf x}
	\leq k_2\int_{\Omega^\epsilon}\av{Du}^p\,d{\bf x}\,,
\end{cases}
}{Teps}
for constants $k_0,\,k_1,\,k_2>0$. Hence, $T_\epsilon$ extends solutions defined
on the pore space \cblu{$\Omega^\epsilon$} to the whole domain $\Omega$.

%% file: equilibrium4.tex
Our main result, i.e., the upscaling/homogenization of general phase field equations (including
the Cahn-Hilliard equation), is based on the following local 
property of the chemical potential.

\begin{defn}\label{def:LoEq} \emph{(Local Thermodynamic Equilibrium)} 
\cblu{Let 
$
\mu(\phi)
	= -\lambda\Delta\phi+\frac{1}{\lambda}f(\phi)
$ 
be the chemical potential associated to the phase field free energy 
density \reff{FrEn}.}
We say that the \cblu{upscaled} chemical potential \cblu{
$
\mu_0(\phi_0) 
	= -\lambda{\rm div}(\hat{\rm D}\nabla\phi_0) + \frac{1}{\lambda}f (\phi_0)
	= \lambda w_0 +\frac{1}{\lambda} f(\phi_0)
$}
is in \emph{local thermodynamic 
equilibrium (LTE)} if and only if 
\bsplitl{
\frac{\partial \mu_0(\phi_0({\bf x}))}{\partial x_k}
	= \begin{cases}
		0 & \text{\cblu{appearing in the cell problem depending on $\Omega\times Y$}}\,,
\\
		\frac{\partial \mu_0(\phi_0)}{\partial x_k} & \text{\cblu{appearing on the macroscale $\Omega$ (after averaging over $Y$)}}\,,
	\end{cases}
}{SSCP}
where $\phi_0(x)$ is the upscaled/slow variable, \cblu{which is independent of 
the microscale ${\bf y}\in Y$ and which} solves the upscaled phase 
field equation (\ref{pmWrThm}) below.
\end{defn}

\begin{rem}\label{rem:LoThEq} Definition \ref{def:LoEq} systematically accounts for the problem specific slow (macroscopic) 
scale ${\bf x} \in\Omega$ and the fast (microscopic) scale ${\bf y}\in Y$. 
\cblu{Intuitively, Definition~\ref{def:LoEq} expresses the fact that 
the macroscopic variables are varying so slowly that their variations are not 
visible on the microscale.} 
The local thermodynamic equilibrium characterization \reff{SSCP} is well 
accepted and appears in a wide range of applications, e.g. \cite{Boda2012,deGroot1969,Malgaretti2013,Kosinska2008}. \hfill$\diamond$
\end{rem}

Definition \ref{def:LoEq} naturally appears in the upscaling of nonlinear problems and enables 
two essential features: a) The upscaled equations are of the same form as the microscopic formulation; 
b) \reff{SSCP} guarantees the well-posedness of arising cell problems which define effective transport coefficients. Recent 
examples in the context of ionic transport equations are \cite{Schmuck2012,Schmuck2013,Schmuck2012a,Schmuck2015}. These considerations allow us to \cblu{recall the following upscaling result from \cite{Schmuck2014}}.

%% file: mainresults7.tex
\medskip

{\bf Upscaling Result (UR):} (Effective macroscopic phase field equations) \emph{
Suppose that $\psi({\bf x})\in H^2_E(\Omega)$. For chemical potentials 
$\mu:=\nabla_\phi E(\phi)$, where $\nabla_\phi$ denotes the Fr\'echet derivative, being in local thermodynamic equilibrium as 
characterized by \emph{Definition \ref{def:LoEq}}, the 
microscopic porous media formulation \reff{PeMoPr} can be effectively
approximated by the following macroscopic problem,
\bsplitl{
\begin{cases}
\quad\theta_1\pd{\phi_0}{t}
    = 
	{\rm div}\brkts{
	\hat{\rm M}_\phi/\lambda\nabla f(\phi_0)
    }
    +\frac{\lambda}{\theta_1}{\rm div}\brkts{
        \hat{\rm M}_w\nabla \brkts{
            {\rm div}\brkts{
                \hat{\rm D}\nabla \phi_0
            }
        }
    }
    &\textrm{in }\Omega_T\,,
\\\quad
\nabla_n \phi_0
    = {\bf n}\cdot\nabla\phi_0
    = 0
    &\textrm{on }\partial\Omega_T\,,
\\\quad
\nabla_n\Delta \phi_0
    = 0
    &\textrm{on }\partial\Omega_T\,,
\\\quad
\phi_0({\bf x},0)
    = \psi({\bf x})
    &\textrm{in }\Omega\,,
\end{cases}
}{pmWrThm}
where $\theta_1:=\frac{\av{Y^1}}{\av{Y}}$ is the porosity and the porous media correction tensors $\hat{\rm D}:=\brcs{{\rm d}_{ik}}_{1\leq i,k\leq d}$,
$\hat{\rm M}_\phi = \brcs{{\rm m}^\phi_{ik}}_{1\leq i,k\leq d}$ and
$\hat{\rm M}_w=\brcs{\cblu{{\rm m}^w_{ik}}}_{1\leq i,k\leq d}$
 are defined by
\bsplitl{
\begin{cases}
\quad
{\rm d}_{ik}
    & := \frac{1}{\av{Y}}\sum^d_{j=1}\int_{Y^1}\brkts{
        \delta_{ik} - \delta_{ij}\pd{\xi^k_\phi}{y_j}
        }
    \,d{\bf y}\,,
\\\quad
{\rm m}^\phi_{ik}
    & :=
    \frac{1}{\av{Y}}\sum_{j=1}^d\int_{Y^1}\brkts{
        {\rm m}_{ik}
        -{\rm m}_{ij}\pd{\xi^k_\phi}{y_j}
    }\,d{\bf y}\,,
\\\quad
{\rm m}^w_{ik}({\bf x})
    & :=
    \frac{1}{\av{Y}}\sum_{j=1}^d\int_{Y^1}\brkts{
        {\rm m}_{ik}
        -{\rm m}_{ij}\pd{\xi^k_w}{y_j}
    }\,d{\bf y}\,,
\end{cases}
}{Dik}
\cblu{where $m_{ij}$ are elements of the mobility tensor.} 
The corrector functions $\xi^k_\phi\in H^1_{per}(Y^1)$ and $\xi^k_w\in L^2(\Omega;H^1_{per}(Y^1))$ for $1\leq k\leq d$
solve in the distributional sense the following reference cell problems
\bsplitl{
\xi_w^k:\,\,
\begin{cases}
-\sum_{i,j,k=1}^d
    \pd{}{y_i}\brkts{
        {\rm m}_{ik}-{\rm m}_{ij}\pd{\xi^k_w}{y_j}
    }
\\\qquad\qquad
    =
    -\sum_{k,i,j=1}^d\pd{}{y_i}\brkts{
        {\rm m}_{ik}
        -{\rm m}_{ij}\pd{\xi^k_\phi}{y_j}
    }
    &\textrm{ in }Y^1\,,
\\
\sum_{i,j,k=1}^d{\rm n}_i\Bigl(
        \brkts{
        {\rm m}_{ij}\pd{\xi^k_w}{y_j}
        -{\rm m}_{ik}
        }
\\\qquad\qquad
	+\brkts{
            {\rm m}_{ik}
            -{\rm m}_{ij}\pd{\xi^k_\phi}{y_j}
        }
    \Bigr)
    = 0
    &\textrm{ on }I_Y:=\partial Y^1\cap\partial Y^2\,,
\\
\xi^k_w({\bf y})\textrm{ is $Y$-periodic and ${\cg M}_{Y^1}(\xi^k_w)=0$,}
\end{cases}
\\
\xi_\phi^k:\,\,
\begin{cases}
-\sum_{i,j=1}^d
    \pd{}{y_i}\brkts{
        \delta_{ik}-\delta_{ij}\pd{\xi^k_\phi}{y_j}
    }
    = 0
    &\textrm{ in }Y^1\,,
\\
\sum_{i,j=1}^d{\rm n}_i
	\brkts{
        \delta_{ij}\pd{\xi^k_\phi}{y_j}
        -\delta_{ik}
        }
    =
    0
    &\textrm{ on }I_Y\,,
\\
\xi^k_\phi({\bf y})\textrm{ is $Y$-periodic and ${\cg M}_{Y^1}(\xi^k_\phi)=0$,}
\end{cases}
}{pmRCTh}
\cblu{where $\delta_{ij}$ is the Kronecker delta function and ${\rm n}_i$ denotes 
the $i$-th component of the outward normal vector $\bf n$.}
}

\begin{figure}
\includegraphics[width=0.345\textwidth]{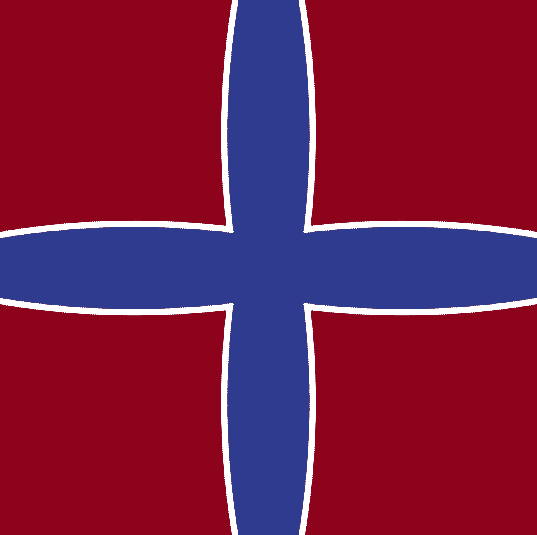}
\hfill
\includegraphics[width=0.57\textwidth]{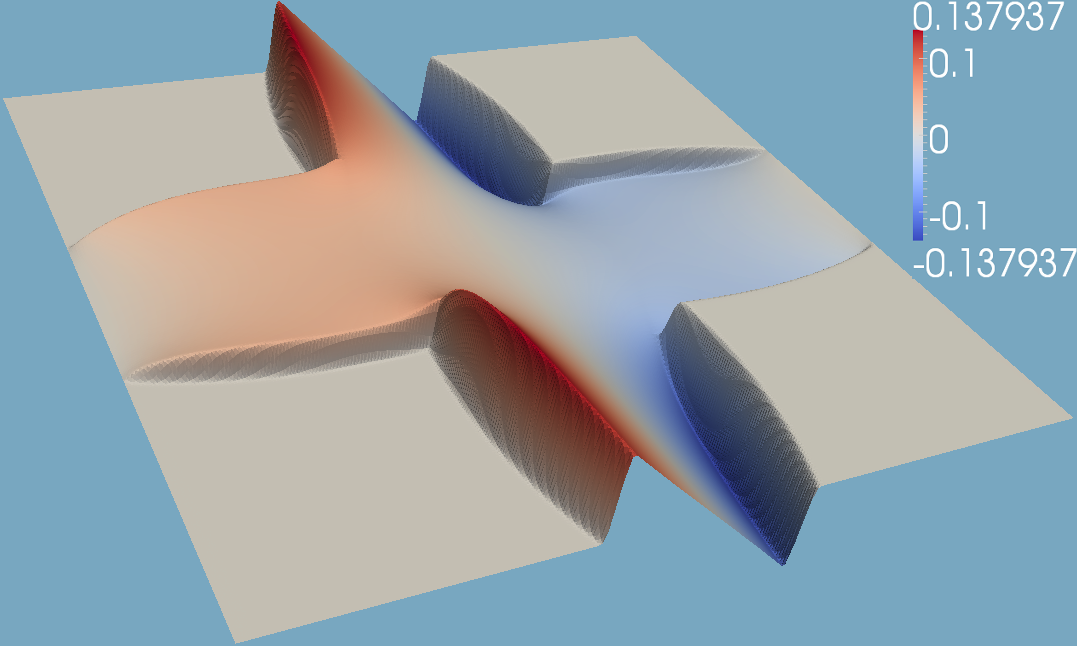}
\caption{{\bf Left:} Pore geometry (shamrock) defined on reference cell (pore phase=blue,solid phase = red). 
{\bf Right:} Corrector $\xi^1_\phi$ solving the cell problem \reff{pmRCTh}$_2$ 
for $Y^1$ representing the pore phase (blue color in the left picture).
}
\label{fig:RefCell}
\end{figure}

\begin{rem}\label{rem:Thm}
i) For an isotropic mobility, i.e., $\hat{\rm M}:= m\hat{\rm I}$ where $\hat{\rm I}$ is the identity matrix, it follows that $\xi^k_w=\xi^k_\phi$, 
and hence both $\xi_w^k$ and $\xi_\phi^k$ solve classical elliptic 
cell problems, see Figure \ref{fig:RefCell}.
\\
ii) The local thermodynamic equilibrium property \emph{(Definition \ref{def:LoEq})} of the macroscopic chemical potential $\mu_0$ enables the derivation of the well-posed cell problem \reff{pmRCTh}$_1$ for $\xi_w^k$. \hfill$\diamond$
\end{rem}

The next result characterizes qualitatively the homogenized phase field equations \reff{pmWrThm} 
with the help of error estimates.

\begin{thm}\label{thm:ErEs} \emph{(Error estimates)} Let $\phi^\epsilon$ 
be a solution of \reff{PeMoPr}, or equivalently $\phi^\epsilon$ and $w^\epsilon$ solve the splitting formulation \reff{DePhMo0}. Suppose that 
\emph{Assumption C} holds.
Moreover, the domain boundaries $\partial\Omega^\epsilon$ and interfaces 
$I^\epsilon_\Omega:=\partial\Omega^\epsilon\cap\partial B^\epsilon$ shall 
be Lipschitz\footnote{see \cite{Chechkin2007} for instance}. Let $\hat{\rm M}=m\hat{\rm I}$ be an isotropic mobility with $\hat{\rm I}$ representing the identity matrix.
If the free energy $F$ is polynomial of \emph{class (PC)}, then the error variables
%
${\rm E}^\phi_\epsilon
	:= \phi^\epsilon-(\phi_0+\epsilon\phi_1)
	\,,
{\rm E}^w_\epsilon
	:= w^\epsilon-(w_0+\epsilon w_1)
$, 
%
%
where 
$w_1:=-\sum_{k=1}^d \xi^k_w({\bf y})\pd{w_0}{x_k}({\bf x},t)$ 
and 
$\phi_1:=-\sum_{k=1}^d \xi^k_\phi({\bf y})\pd{\phi_0}{x_k}({\bf x},t)$, 
satisfy for $0\leq t\leq T$ and $0<T<\infty$ the following estimates
\bsplitl{
\N{E_\epsilon^w(\cdot,t)}{L^2(\Omega^\epsilon)}^2
	& +c(m,\lambda,\kappa)\int_0^t\N{{\cal A}_\epsilon E_\epsilon^w(\cdot,s)}{L^2(\Omega^\epsilon)}^2\,ds
\\&	
	\leq \epsilon^{1/2}C(T,\Omega,m,\kappa,\lambda)
	\,,
\\
\N{E_\epsilon^\phi(\cdot,t)}{H^1(\Omega^\epsilon)}
	& \leq \epsilon^{1/4}C(T,\Omega,m,\kappa,\lambda)
	\,,
}{ThmErEs}
where $c(m,\lambda,\kappa)$ and $C(T,\Omega,m,\kappa,\lambda)$ are 
constants independent of $\epsilon$.
\end{thm}

\begin{rem}\label{rm:Rate} We note that the proof of the above 
Theorem \ref{thm:ErEs} does not take the behaviour in the boundary 
region into account by solely applying a smooth enough truncation. 
This leads for linear 
elliptic equations to the by now classical convergence rate $\epsilon^{1/2}$, 
e.g. \cite{Chechkin2007,Zhikov2006}. 
However, in recent attempts \cite{Suslina2012,Pastukhova2013}, the 
authors can improve the convergence rates with the help of operator 
estimates with a resulting rate $\epsilon$. We note, that our estimates in \reff{ThmErEs} are 
derived based on the classical method but due to the fourth order 
operator, we end up with the slightly lower rate $\epsilon^{1/4}$, albeit 
under the generally required strong regularity Assumption C. The strongest regularity 
result currently available seems to be the estimates stated in 
Lemma \ref{lem:ApEs}.
\end{rem}

To the best of our knowledge, this is the first error quantification in terms of convergence 
rates with respect to the heterogeneity $\epsilon$ of the porous media approximation \reff{pmWrThm} for phase field equations. The estimates \reff{ThmErEs} imply convergence of solutions $\phi^\epsilon$ 
of the microscopic formulation \reff{PeMoPr} to solutions $\phi_0$ of the upscaled problem \reff{pmWrThm} 
for a vanishing heterogeneity parameter based on the 
regularity Assumption C.

%% file: upscaling4.tex
For convenience, we first recall here the formal derivation of the effective 
macroscopic phase field equation from \cite{Schmuck2014}. 
For the micro-scale variable $\frac{{\bf x}}{\epsilon}=:{\bf y}\in Y$ it 
holds that, 
$
\pd{f_\epsilon({\bf x})}{x_i}
    = \frac{1}{\epsilon}\pd{f}{y_i}({\bf x}, {\bf x}/\epsilon)
    +\pd{f}{x_i}({\bf x}, {\bf x}/\epsilon)\,,
\textrm{ and }
\nabla f_\epsilon ({\bf x})
    =\cblu{ \frac{1}{\epsilon}\nabla_yf({\bf x},{\bf x}/\epsilon)
    +\nabla_xf({\bf x}, {\bf x}/\epsilon)}\,,
$
\,\,where $f_\epsilon({\bf x})=f({\bf x},{\bf  y})$ is an arbitrary function depending on two variables ${\bf x}\in\Omega$,
${\bf y}\in Y$. Hence, we have
\bsplitl{
\begin{cases}
\quad
{\cg A}_0
   & =
    -\sum_{i,j=1}^d\pd{}{y_i}\brkts{\delta_{ij} \pd{}{y_j}}
    \,,
\\\quad
{\cg A}_1
    &=
    -\sum_{i,j=1}^d\biggl[\pd{}{x_i}\brkts{\delta_{ij}\pd{}{y_j}}
        +\pd{}{y_i}\brkts{\delta_{ij}\pd{}{x_j}}
        \biggr]\,,
\\\quad
{\cg A}_2
   & = - \sum_{i,j=1}^d\pd{}{x_j}\brkts{\delta_{ij}\pd{}{x_j}}\,,
\\\quad
{\cg B}_0
   & =
    - \sum_{i,j=1}^d\pd{}{y_i}\brkts{{\rm m}_{ij}\pd{}{y_j}}
    \,,
\\\quad
{\cg B}_1
   & =
    -\sum_{i,j=1}^d\biggl[\pd{}{x_i}\brkts{{\rm m}_{ij}\pd{}{y_j}}
        +\pd{}{y_i}\brkts{{\rm m}_{ij}\pd{}{x_j}}
        \biggr]\,,    
\\\quad
{\cg B}_2
    & = - \sum_{i,j=1}^d\pd{}{x_j}\brkts{{\rm m}_{ij}\pd{}{x_j}}
    \,.
\end{cases}
}{B0B1B2}
Herewith, we can define ${\cg A}_\epsilon := \epsilon^{-2}{\cg A}_0
+ \epsilon^{-1}{\cg A}_1 +{\cg A}_2$ and analogously ${\cg B}_\epsilon$. Hence, it holds for the
Laplace operator that
$\cblu{-\Delta f_\epsilon({\bf x}) =
{\cg A}_\epsilon f({\bf x}, {\bf y})}$. 
In order to deal with the multiscale nature of strongly heterogeneous 
environments \cite{Hornung1997,Schmuck2012,Schmuck2012a,Schmuck2013}, 
the following, formal asymptotic expansions are used,
\bsplitl{
\zeta^\epsilon
    & \approx \zeta_0({\bf x},{\bf y},t)
    +\epsilon \zeta_1({\bf x},{\bf y},t)
    +\epsilon^2 \zeta_2({\bf x},{\bf y},t)\,,
\quad
\textrm{for}\quad\zeta\in\brcs{w,\phi}\,,
}{AsExp}
where higher order terms are neglected. 
Before we can insert \reff{AsExp} into the microscopic formulation \reff{DePhMo0}, we need to approximate the derivative of the 
nonlinear homogeneous free energy $f:=F'$ by a Taylor expansion of the form
\bsplitl{
	f(\phi^\epsilon)
	\approx
	f(\phi_0)+ f'(\phi_0)(\phi^\epsilon-\phi_0)
	+ {\cal O}\brkts{(\phi^\epsilon-\phi_0)^2}\,,
}{FTE}
where $\phi_0$ stands for the leading order term in \reff{AsExp}.\footnote{
Here, we allow for general free energy densities in difference to the subsequent rigorous derivation 
of error estimates which is based on energy densities of the polynomial class (PC).
}
Using \reff{AsExp} and \reff{FTE} in \reff{DePhMo0} \cblu{with 
$\nabla_n\phi=g$} and with \reff{B0B1B2}, 
we get the following sequence of problems,
\bsplitl{
\mathcal{O}(\epsilon^{-2}):\quad
\begin{cases}
\mathcal{B}_0\ebrkts{
		\lambda w_0
		+1/\lambda f(\phi_0)
	}
    = 0
    &\textrm{in }Y^1\,,
\\\quad
\textrm{no flux b.c.}\,,
\\\quad
\textrm{$w_0$ is $Y^1$-periodic}\,,
\\
\mathcal{A}_0\phi_0=0
    &\textrm{in }Y^1\,,
\\\quad
\nabla_n \phi_0
    = 0
    &\textrm{on }\partial Y^1\cap\partial Y^2\,,
\\\quad
\textrm{$\phi_0$ is $Y^1$-periodic}\,,
\end{cases}
}{O-2n}
\bsplitl{
\mathcal{O}(\epsilon^{-1}):\hspace{10cm}
\\
\begin{cases}
{\cg B}_0 
	\ebrkts{
		\lambda w_1
		+1/\lambda f'(\phi_0)\phi_1
	}
    = -{\cg B}_1\ebrkts{ 
		\lambda w_0
		+1/\lambda f(\phi_0)
	}
    &\textrm{in }Y^1\,,
\\\quad
\textrm{no flux b.c.}\,,
\\\quad
\textrm{$w_1$ is $Y^1$-periodic}\,,
\\
{\cg A}_0 \phi_1
    = -{\cg A}_1 \phi_0
    &\textrm{in }Y^1\,,
\\\quad
\nabla_n \phi_1
    = 0
    &\textrm{on }\partial Y^1\cap\partial Y^2\,,
\\\quad
\textrm{$\phi_1$ is $Y^1$-periodic}\,,
\end{cases}
}{O-1n}
\bsplitl{
\mathcal{O}(\epsilon^{0}):\hspace{10.5cm}
\\
\begin{cases}
{\cg B}_0 \ebrkts{
		\lambda w_2
		+\frac{1}{\lambda}\brkts{
			\frac{1}{2}f''(\phi_0)\phi_1^2+f'(\phi_0)\phi_2
		}
	}
	&
\\ \qquad
    = -\brkts{
        {\cg B}_2\ebrkts{
		 \lambda w_0
		+1/\lambda f(\phi_0)
	}
        +{\cg B}_1\ebrkts{
		\lambda w_1
		\cblu{+}1/\lambda f'(\phi_0)\phi_1
	}
	}
	&
\\ \qquad\qquad\qquad\quad
    -\partial_t\left(-\Delta\right)^{-1} w_0
    &\textrm{in }Y^1\,,
\\\quad
\textrm{no flux b.c.}\,,
\\\quad
\textrm{$w_2$ is $Y^1$-periodic}\,,
\\
{\cg A}_0 \phi_2
    = -{\cg A}_2 \phi_0
    -{\cg A}_1 \phi_1
    +w_0
    &\textrm{in }Y^1\,,
\\\quad
\nabla_n \phi_2
    = g_\epsilon
    &\textrm{on }\partial Y^1\cap\partial Y^2\,,
\\\quad
\textrm{$\phi_2$ is $Y^1$-periodic}\,.
\end{cases}
}{O-0n}

Problem \reff{O-2n} immediately suggests, based on classical homogenization theory \cite{Bensoussans1978}, that $\phi_0$ is independent of the microscale ${\bf
y}$. This and the linear structure of \reff{O-1n} allow for the
following ansatz for $w_1$ and $\phi_1$, i.e., 
\bsplitl{
w_1({\bf x},{\bf y},t)
    & = -\sum_{k=1}^d \xi^k_w({\bf y})\pd{w_0}{x_k}({\bf x},t)\,,
\quad
\phi_1({\bf x},{\bf y},t)
    = -\sum_{k=1}^d \xi^k_\phi({\bf y})\pd{\phi_0}{x_k}({\bf x},t)
    \,.
}{XiwXiphi}

Plugging \reff{XiwXiphi} into \reff{O-1n}$_2$ gives an equation for 
$\xi^k_w$ and \cblu{$\xi^k_\phi$}. 
The resulting equation for \cblu{$\xi^k_\phi$} can be immediately written
for $1\leq k\leq d$ as,
\bsplitl{
\xi_\phi:\quad
\begin{cases}
-\sum_{i,j=1}^d
    \pd{}{y_i}\brkts{
        \delta_{ik}-\delta_{ij}\pd{\xi^k_\phi}{y_j}
    }
    =
    &
\\\qquad\quad\,\,
    = -{\rm div}\brkts{
        {\bf e}_k-\nabla_y\xi^k_\phi
    }=0
    &\textrm{ in }Y^1\,,
\\
        {\bf n}\cdot\brkts{
            \nabla\xi^k_\phi
            +{\bf e}_k
        }
    =
    0
    &\textrm{ on }\partial Y^1\cap\partial Y^2\,,
\\
\xi^k_\phi({\bf y})\textrm{ is $Y$-periodic and ${\cg M}_{Y^1}(\xi^k_\phi)=0$.}
\end{cases}
}{Xiphi}

To study \reff{O-1n}$_1$, we first rewrite ${\cg B}_0\ebrkts{f'(\phi_0)\phi_1}$ and ${\cg B}_1f(\phi_0)$ as follows
\bsplitl{
{\cg B}_0\ebrkts{f'(\phi_0)\phi_1}
	& = -\sum_{k,i,j=1}^d\frac{\partial}{\partial y_i}\brkts{
		{\rm m}_{ij}\frac{\partial\xi^k_\phi}{\partial y_j}\frac{\partial f(\phi_0)}{\partial x_k}
	}\,,
\\
{\cg B}_1 f(\phi_0)
	& = \cblu{-}\sum_{i,j=1}^d
		\frac{\partial}{\partial y_i}\brkts{
			{\rm m}_{ij}
			\frac{\partial f(\phi_0)}{\partial x_j}
		}\,.
}{O-1xpl}
\cblu{Rewriting $w_1$ and $w_0$ in the same way} and using \reff{XiwXiphi} leads then to 
\bsplitl{
-\lambda\sum_{k,i,j=1}^d
	\frac{\partial}{\partial y_i}
	&\brkts{
		{\rm m}_{ij}\brkts{
			\frac{\partial x_k}{\partial x_j}
			-\frac{\partial\xi^k_w}{\partial y_j}
		}\frac{\partial w_0}{\partial x_k}
	}
\\&
	= 1/\lambda\sum_{k,i,j=1}^d
	\frac{\partial}{\partial y_i}\brkts{
		{\rm m}_{ij}\brkts{
			\frac{\partial x_k}{\partial x_j}
			-\frac{\partial\xi^k_\phi}{\partial y_j}
		}\frac{\partial f(\phi_0)}{\partial x_k}
	}\,,
}{O-1n1}
in $Y^1$. Next, \cblu{due to local thermodynamic equilibrium property of 
the upscaled chemical potential $\mu_0(\phi_0)$ as defined in Definition~\ref{def:LoEq}}, we have on the level of the 
reference cell $Y$,
\bsplitl{
\cblu{\frac{\partial\mu_0}{\partial x_i}
	=
	\frac{\partial}{\partial x_i}\brkts{
		f(\phi_0)/\lambda
		-\lambda{\rm div}(\hat{\rm D}\nabla\phi_0)
	}
	=
	\frac{\partial}{\partial x_i}\brkts{
		f(\phi_0)/\lambda
		+\lambda w_0
	}
	}
	= 0\,.
}{sclsep}
Entering with \reff{sclsep} into \reff{O-1n1} finally gives the reference cell problem for 
$\xi^k_w$, $1\leq k\leq d$ for given \cblu{$\xi^k_\phi$}
\bsplitl{
\begin{cases}
-\sum_{i,j=1}^d
    \pd{}{y_i}\brkts{
        {\rm m}_{ik}-{\rm m}_{ij}\pd{\xi^k_w}{y_j}
    }
\\\qquad\qquad
    =
    -\sum_{i,j=1}^d\pd{}{y_i}\brkts{
        {\rm m}_{ik}
        -{\rm m}_{ij}\pd{\xi^k_\phi}{y_j}
    }
    &\textrm{ in }Y^1\,,
\\
\sum_{i,j=1}^d{\rm n}_i\Bigl(
        \brkts{
        {\rm m}_{ij}\pd{\xi^k_w}{y_j}
        -{\rm m}_{ik}
        }
	+\brkts{
            {\rm m}_{ik}
            -{\rm m}_{ij}\pd{\xi^k_\phi}{y_j}
        }
    \Bigr)
    = 0
    &\textrm{ on }\partial Y^1\cap\partial Y^2\,,
\\
\xi^k_w({\bf y})\textrm{ is $Y$-periodic and ${\cg M}_{Y^1}(\xi^k_w)=0$,}
\end{cases}
}{Xiw}

Finally, we consider the last problem \reff{O-0n}. 
Standard existence and uniqueness results
(Fredholm alternative/Lax-Milgram) guarantee solvability
after validating that the right hand side in \reff{O-0n} is zero as an
integral over $Y^1$. This means, 
$-\sum_{i,j=1}^d\int_{Y^1}
    \pd{}{x_i}\brkts{
        \delta_{ij}\brkts{
            \pd{\phi_1}{y_j}
            +\pd{\phi_0}{x_j}
        }
    }\,d{\bf y}
    -\tilde{g}_0
    =\av{Y^1}w_0
$\,,
where 
$$
\tilde{g}_0:=-\frac{\gamma}{C_h}\int_{\partial Y^1}\brkts{a_1\chi_{\partial Y^1_{w_1}}
    +a_2\chi_{\partial Y^1_{w_2}}}\,d{\bf y}
	\,,
$$ 
which leads to
\bsplitl{
-\sum_{i,k=1}^d\ebrkts{
        \sum_{j=1}^d\int_{Y^1}\brkts{
            \delta_{ik}-\delta_{ij}\pd{\xi^k_\phi}{y_j}
        }\,d{\bf y}
    }\pd{^2\phi_0}{x_i\partial x_k}
    = \av{Y^1}w_0
    +\tilde{g}_0
    \,.
}{UpPhi1}
\cblu{The inhomogeneous Neumann boundary condition $\tilde{g}_0$ 
accounts for pore walls $\partial Y^1_{w_1}\cup\partial Y^1_{w_1}=\partial Y^1$ showing two specific wetting properties 
characterised by the parameters $a_1$ and $a_2$ specifying the walls $\partial Y^1_{w_1}$ and $\partial Y^1_{w_2}$, respectively. 
The upscaling result \reff{pmWrThm} is stated for neutral wetting 
characteristics of the pore walls, i.e., $\tilde{g}=0$. 
} 
\reff{UpPhi1} suggests to define a porous media correction tensor
$\hat{\rm D}:=\brcs{{\rm d}_{ik}}_{1\leq i,k\leq d}$ by
\bsplitl{
\av{Y}{\rm d}_{ik}
    := \sum_{j=1}^d\int_{Y^1}\brkts{
        \delta_{ik}
        -\delta_{ij}\pd{\xi^k_\phi}{y_j}
    }\,d{\bf y}\,.
}{Dphi}
Equations \reff{UpPhi1} and \reff{Dphi} represent the upscaled
equation for $\phi_0$, i.e., $-\Delta_{\hat{\rm D}} \phi_0
    :=
    -{\rm div}\brkts{
        \hat{\rm D}\nabla \phi_0
    }
    = \theta_1w_0
    +\frac{1}{|Y|}\tilde{g}_0$.

The upscaled equation for $w$ is again a result of the Fredholm alternative, i.e., a solvability
criterion on equation \reff{O-0n}$_1$. We require,
\bsplitl{
\int_{Y^1}\Bigl\{
    -\lambda\brkts{
         {\cg B}_2 w_0
        +{\cg B}_1 w_1
    }
    -\frac{1}{\lambda}{\cg B}_1\ebrkts{
        f'(\phi_0)\phi_1
    }
    -\frac{1}{\lambda}{\cg B}_2f(\phi_0)
    -\partial_t{\cal A}_2^{-1}w_0
    \Bigr\}\,d{\bf y}
    =0\,.
}{SoCow2}
The first two terms in \reff{SoCow2} can be rewritten by,
\bsplitl{
\int_{Y^1}-\brkts{
        {\cg B}_2 w_0
        +{\cg B}_1 w_1
    }\,d{\bf y}
    =
    \sum_{i,k=1}^d\ebrkts{
        \sum_{j=1}^d\int_{Y^1}\brkts{
            {\rm m}_{ik}-{\rm m}_{ij}\pd{\xi^k_w}{y_j}
        }\,d{\bf y}
    }\pd{^2w_0}{x_i\partial x_k}
\\
    = {\rm div}\brkts{
        \hat{\rm M}_w\nabla w_0
    }
    \,,
}{Bw}
where the effective tensor $\hat{\rm M}_w=\brcs{{\rm m}^w_{ik}}_{1\leq i,k\leq d}$ is defined by
$${\rm m}^w_{ik}
    :=
    \frac{1}{\av{Y}}\sum_{j=1}^d\int_{Y^1}\brkts{
        {\rm m}_{ik}
        -{\rm m}_{ij}\pd{\xi^k_w}{y_j}
    }\,d{\bf y}\,.
$$
The third term in \reff{SoCow2} becomes
\bsplitl{
\hspace{-0.1cm}
-{\cg B}_1
	&\ebrkts{
		f'(\phi_0)\phi_1
    	}
	=
\\&\hspace{-0.75cm}	
	-\sum_{i,j=1}^d\ebrkts{
		\frac{\partial}{\partial x_i}\brkts{
			{\rm m}_{ij}f'(\phi_0)
			\sum_{k=1}^d\frac{\partial\xi^k_\phi}{\partial y_j}\frac{\partial\phi_0}{\partial x_k}
		}
		+\frac{\partial}{\partial y_i}\brkts{
			{\rm m}_{ij}f'(\phi_0)\sum_{k=1}^d\xi^k_\phi\frac{\partial^2\phi_0}{\partial x_k\partial x_j}
		}
	},
}{Bphi}
where the last term in \reff{Bphi} disappears after integrating by parts. The first term 
on the right-hand side of \reff{Bphi} can be rewritten with the help of the chain rule
$\frac{\partial^2 f(\phi_0)}{\partial x_k\partial x_j}
	= f''(\phi_0)\frac{\partial\phi_0}{\partial x_k}\frac{\partial\phi_0}{\partial x_j}
	+ f'(\phi_0)\frac{\partial^2\phi_0}{\partial x_k\partial x_j}
$\,,
as follows
$$-{\cg B}_1
	\ebrkts{
		f'(\phi_0)\phi_1
    	}
	=
	-\sum_{i,j=1}^d {\rm m}_{ij}\sum_{k=1}^d\frac{\partial\xi^k_\phi}{\partial y_j}
	\frac{\partial^2 f(\phi_0)}{\partial x_k\partial x_i}
	\,,
$$
to which we add the term $-{\cal B}_2f(\phi_0)$. Herewith, we define a 
tensor \cblu{$\hat{\rm M}_\phi=\brcs{{\rm m}^\phi_{ij}}_{1\leq i,k\leq d}$} 
by 
$$
{\rm m}^\phi_{ik}
    :=
    \frac{1}{\av{Y}}\sum_{j=1}^d\int_{Y^1}\brkts{
	{\rm m}_{ik}
        -{\rm m}_{ij}\pd{\xi^k_\phi}{y_j}
    }\,d{\bf y}\,,
$$
which allows us to write
\bsplitl{
\cblu{
\int_{Y^1}\Bigl(
	-{\cg B}_1
	\ebrkts{
		f'(\phi_0)\phi_1
    	}
	-{\cg B}_2f(\phi_0)
	\Bigr)
	\, d{\bf y}
	}
	= {\rm div}\brkts{
		\hat{\rm M}_\phi\nabla f(\phi_0)
	}
	\,.
}{B1B2}

\cblu{These considerations together with the identity 
$$
\partial_t {\cal A}_2^{-1}w_0
	=\partial_t {\cal A}_2^{-1}\brkts{{\cal A}_2\phi_0+{\cal A}_1\phi_1}
	=\partial_t\phi_0 +\partial_t{\cal A}_2^{-1}{\cal A}_1\phi_1
	\,,
$$ 
where the last term subsequently disappears due to $Y$-periodicity,} 
finally lead after integration over the microscale $Y$ to
the following effective equation for $\phi_0$, i.e.,
\bsplitl{
\theta_1\pd{\phi_0}{t}
    = {\rm div}\brkts{
	\hat{\rm M}_\phi/\lambda\nabla f(\phi_0)
    }
    +\frac{\lambda}{\theta_1}{\rm div}\brkts{
        \hat{\rm M}_w\nabla \brkts{
            {\rm div}\brkts{
                \hat{\rm D}\nabla \phi_0
            }
            -\tilde{g}_0
        }
    }\,.
}{EfW0}
The solvability of \reff{EfW0} follows along with the arguments in \cite{Novick-Cohen1990} via 
a local Lipschitz argument, or via a Galerkin approximation and 
a priori estimates as developed in \cite[Theorem 4.2, p. 155]{Temam1997}. 

%% file: error.tex
For the derivation of the error estimates \reff{ThmErEs}, we work with the splitting formulation introduced in \cite{Schmuck2012b} and summarized in \reff{DePhMo0}. We extend the derivation of error estimates for 
second order problems \cite[Theorem 6.3, Section 6.2 \& Section 7.2]{Cioranescu1999} and \cite{Schmuck2012} to the fourth order phase field problems studied here. Hence, we compare the solution of the microscopic porous media formulation \reff{PeMoPr} with the solution of the effective upscaled porous media formulation \reff{pmWrThm}. 
As in \cite{Cioranescu1999,Schmuck2012}, we introduce the error variables 
\bsplitl{
{\rm E}^w_\epsilon
	& := w^\epsilon-(w_0+\epsilon w_1)
	\,,
\\
{\rm E}^\phi_\epsilon
	& := \phi^\epsilon-(\phi_0+\epsilon\phi_1)
	\,,
\\ 
\cblu{
{\rm E}^f_\epsilon 
	}
	& \cblu{ := f(\phi^\epsilon)
	-\brkts{
		f(\phi_0)
		+f'(\phi_0)({\rm E}^\phi_\epsilon+\epsilon\phi_1)
	}
	} 
%
	\,,
}{Vars}
here extended by the error function $E^f_\epsilon$.
The first goal is to determine the variables ${\rm F}_\epsilon^\iota$ and ${\rm G}_{\epsilon}^\iota$ which allow us to 
write the equations for the errors ${\rm E}_\epsilon^\iota$ for $\iota\in\brcs{w,\phi}$ as follows
\bsplitl{
\begin{cases}
\quad
\cblu{
\frac{\partial \left(-\Delta\right)^{-1}{\rm E}^w_\epsilon}{\partial t}
	}
	= 
		{\cg B}_\epsilon\ebrkts{ 
			-\lambda{\rm E}^w_\epsilon
			+ 
			\frac{1}{\lambda}\cblu{{\rm E}^f_\epsilon(\phi^\epsilon,\phi_0,\phi_1)}
		}
	+ {\rm R}^\phi_\epsilon
	+ \epsilon {\rm F}^w_\epsilon
	& \textrm{in }\Omega^\epsilon\times ]0,T[
	\,,
\\\qquad\quad
\nabla_n{\rm E}^w_\epsilon
	= \epsilon {\rm G}^w_\epsilon
	& \textrm{on }\partial\Omega^\epsilon\times ]0,T[
	\,,
\\\quad
{\cg A}_\epsilon {\rm E}^\phi_\epsilon 
	= {\rm E}^w_\epsilon
	+ \epsilon {\rm F}^\phi_\epsilon
	& \textrm{in }\Omega^\epsilon\times ]0,T[
	\,,
\\\qquad\quad
\nabla_n{\rm E}^\phi_\epsilon
	= \epsilon {\rm G}^\phi_\epsilon
	& \textrm{on }\partial\Omega^\epsilon\times ]0,T[
	\,.
\end{cases}
}{ErEq}
With the definitions \reff{B0B1B2} we can rewrite the first term on the right-hand side in \reff{ErEq}$_1$ and 
the term on the left-hand side in \cblu{\reff{ErEq}$_3$} as follows
\bsplitl{
	{\cg B}_\epsilon
	\ebrkts{ 
			-\lambda{\rm E}^w_\epsilon
			+ 
			\cblu{\frac{1}{\lambda}{\rm E}^f_\epsilon(\phi^\epsilon,\phi_0,\phi_1)}
		}
	& = \brcs{
		\epsilon^{-2}{\cal B}_0
		+\epsilon^{-1}{\cal B}_1
		+{\cal B}_2
	}\Bigl[
		-\lambda{\rm E}^w_\epsilon
\\&\quad
			+ 
			\cblu{\frac{1}{\lambda}{\rm E}^f_\epsilon(\phi^\epsilon,\phi_0,\phi_1)}
	\Bigr]\,,
\\
{\cg A}_\epsilon {\rm E}_\epsilon^\phi
	& = \brcs{
		\epsilon^{-2}{\cal A}_0
		+\epsilon^{-1}{\cal A}_1
		+{\cal A}_2
	}{\rm E}_\epsilon^\phi\,.
}{Re1TeRHS}
The relations \cblu{\reff{B0B1B2} and} \reff{Re1TeRHS} together with \cblu{the sequence of problems \reff{O-2n}, \reff{O-1n}, and \reff{O-0n} define the terms 
\cblu{${\rm R}^\phi_\epsilon$}, ${\rm F}^w_\epsilon$ and ${\rm F}^\phi_\epsilon$} by
\bsplitl{
\cblu{
{\rm R}^\phi_\epsilon
	}
	&\cblu{ 
	:= 
	-m\Delta\brkts{
		\frac{1}{\lambda}f'(\phi_0){\rm E}^\phi_\epsilon
	}
	}
	\,,
\\
{\rm F}^w_\epsilon
	& := \cblu{
	{\cal B}_2\brkts{ 
				\lambda w_1
				-\frac{1}{\lambda}f'(\phi_0)\phi_1
	}
	+\frac{\partial}{\partial t}(-\Delta)^{-1}w_1
	}
	\,,
\\
{\rm F}^\phi_\epsilon
	& := -\brkts{
			{\cg A}_2 \phi_1
			\cblu{+w_1}
		}
		\,,
}{Fweps}
\cblu{since ${\cal B}_\epsilon=-m\Delta$. The definitions in \reff{Fweps} 
are a consequence of the two identities
\bsplitl{
\brcs{
\frac{\partial}{\partial t}(-\Delta)^{-1}w^\epsilon
	+{\cal B}_\epsilon\brkts{
			\lambda w_\epsilon
			-\frac{1}{\lambda} f(\phi_\epsilon)
		}
	}
	& = 
	\biggl\{
		\frac{\partial}{\partial t}(-\Delta)^{-1}w_0
		+{\cal B}_2\brkts{
			\lambda w_0
			-\frac{1}{\lambda}f(\phi_0)
		}
\\&
		+{\cal B}_1\brkts{
			\lambda w_1
			-\frac{1}{\lambda}f'(\phi_0)\phi_1
		}
	\biggr\}
	+ {\rm R}^\phi_\epsilon
	+ \epsilon{\rm F}^w_\epsilon
	\,,
\\
\brcs{
		{\cal A}_\epsilon {\rm E}^\phi_\epsilon
		- {\rm E}^w_\epsilon
	}
	= 
	\epsilon{\rm F}^\phi_\epsilon
	\,, 
}{DrErrEq}
where the terms in the braces vanish due to the microscopic and 
homogenized equations and these terms represent the first and the 
second term in the error equation \reff{ErEq}$_1$ and 
\reff{ErEq}$_3$. 
} 
The inhomogeneities in the boundary conditions in \reff{ErEq} satisfy
\bsplitl{
{\rm G}^w_\epsilon
	& 
	:= -\cblu{\nabla_n} w_1 
	= \cblu{\nabla_n}\sum_{k=1}^d\xi_w^k\frac{\partial w_0}{\partial x_k}
	\,,
\quad\textrm{and}\\
{\rm G}^\phi_\epsilon 
	&
	:= -\cblu{\nabla_n}\phi_1 
	= \cblu{\nabla_n}\sum_{k=1}^d\xi_\phi^k\frac{\partial \phi_0}{\partial x_k}
	\,,
}{GwGphi}
since the boundary conditions imply $\nabla_n(\iota^\epsilon-\iota_0)=0$ for 
$\iota\in\brcs{w,\phi}$. 
Under the given regularity (Assumption C) of boundaries and interfaces, i.e., $\partial\Omega\in C^\infty$ 
as well as $I_Y\in C^\infty$, we obtain that $\xi_\phi^k,\,\xi_w^k\in W^{1,\infty}$ by classical regularity theory for elliptic problems \cite{Gilbarg2001}. 
We note that the above regularity requirements on $\partial\Omega$ and 
$I_Y$ are not necessarily sharp but this question is beyond the scope of 
this work. 
Hence, elliptic theory allows us also to estimate \reff{ErEq}$_3$ by
\bsplitl{
\N{{\rm E}^\phi_\epsilon}{H^1(\Omega^\epsilon)}
	& \leq 
		C\brkts{
			\N{{\rm E}^w_\epsilon}{L^2(\Omega^\epsilon)}
			+\epsilon \N{{\rm F}^\phi_\epsilon}{L^2(\Omega^\epsilon)}
			+\epsilon\N{G^\phi_\epsilon}{H^{-1/2}(\partial\Omega^\epsilon)}
		}
\\&
	\leq 
		C\N{{\rm E}^w_\epsilon}{L^2(\Omega^\epsilon)}
		+\epsilon C\brkts{1+\epsilon^{-1/2}}
	\,,
}{ClElEs}
where we subsequently justify the uniform boundedness of ${\rm F}^\phi_\epsilon$ 
in $L^2(\Omega^\epsilon)$ \cblu{and the $\epsilon$-dependent bound on} ${\rm G}_\epsilon^\phi\in H^{1/2}(\partial\Omega^\epsilon)$.

\cblu{Next, we derive bounds for the terms on the right-hand side, i.e., 
\reff{Fweps}.}
Thanks to Assumption C, it holds that
\bsplitl{
\N{{\rm F}^\phi_\epsilon}{L^2(\Omega^\epsilon)}
	\leq \cblu{C\sum_{i,k,l=1}^d\N{\frac{\partial^3\phi_0}{\partial x_i\partial x_k\partial x_l}}{L^\infty(\Omega^\epsilon)}
			\N{\delta_{ik}\xi^k_\phi\brkts{\frac{\cdot}{\epsilon}}}{L^2(\Omega^\epsilon)}
			+\N{w_1}{L^2(\Omega^\epsilon)}
	}
	\leq C\,,
}{FphiL2}
for a constant $C>0$ independent of $\epsilon$. Analogously, we obtain the bound
\bsplitl{
&\N{{\rm F}^w_\epsilon}{L^2(\Omega^\epsilon)}
	\leq \cblu{
		C\sum_{i,k,l=1}^d\N{\frac{\partial^3w_0}{\partial x_i\partial x_k\partial x_l}}{L^\infty(\Omega^\epsilon)}
			\N{\delta_{ik}\xi^k_w\brkts{\frac{\cdot}{\epsilon}}}{L^2(\Omega^\epsilon)}
		+\frac{1}{\lambda}
				\Ll{
					{\cg B}_2f'(\phi_0)\phi_1
				}
	}
\\&\qquad\qquad\qquad
	\cblu{
				+\N{\frac{\partial}{\partial t}(-\Delta)^{-1}w_1}{L^2(\Omega^\epsilon)}
	}
		\leq C
		\,.
}{FwL2}
We show the basic steps to bound the second last term in \reff{FwL2}. 
To this end, we first note that 
${\cal B}_2 [f'(\phi_0)\phi_1]=-m\Delta_x[f'(\phi_0)(\pmb{\xi}_\phi\cdot\nabla_x)\phi_0]$ 
from which we can identify the 
three most challenging terms to estimate:
\bsplitl{
\begin{cases}
\quad
-\Delta_x f'(\phi_0)(\pmb{\xi}_\phi\cdot\nabla_x)\phi_0
	\,,
	&\quad
\\
\quad
	-f'(\phi_0)(\pmb{\xi}_\phi\cdot\nabla_x)\Delta_x\phi_0
	\,,
	&\quad
\\
\quad
	-f'(\phi_0)(\Delta_x\pmb{\xi}_\phi\cdot\nabla_x)\phi_0
	\,,&
\end{cases}
}{3cht}
\cblu{where ${\bf \xi}_\phi$ is the vector consisting of the 
corrector elements $\xi_\phi^k$ defined in \reff{Xiphi}.} 
In order to bound the terms containing $\phi_0$ we use Assumption C. 
The factor $f'(\phi_0)$ and associated derivatives in \reff{3cht} 
can be bounded by the fact that $f(s)$ is a polynomial of order $2p-1$ 
with characterization (PC), i.e.,
\bsplitl{
\begin{cases}
\quad
\av{f'(s)}
	&\leq c\brkts{
		1+\av{s}^{2p-2}
	}
	\,,
\\
\quad
\av{f''(s)}
	& \leq c\brkts{
		1+\av{s}^{2p-3}
	}
	\,,
\\
\quad
\av{f'''(s)}
	& \leq c\brkts{
		1+\av{s}^{2p-4}
	}
	\,.
\end{cases}
}{Polyf}
We also note that $-\Delta_x\xi^k_\phi=-1/\epsilon^2\Delta_y\xi^k_\phi=0$ 
in view of the cell/corrector problem for $\xi^k_\phi$, $1\leq k\leq d$. 

\cblu{
Finally, the remaining term ${\rm R}_\epsilon^\phi$ first 
decomposes in view of Assumption C 
as follows,
\bsplitl{
\av{{\rm R}_\epsilon^\phi}
	& = 
	\av{
	\frac{m}{\lambda}\brkts{
		{\rm div}\brkts{
			f'(\phi_0)\nabla {\rm E}^\phi_\epsilon
		}
		+
		{\rm div}\brkts{
			f''(\phi_0)\nabla\phi_0{\rm E}^\phi_\epsilon
		}
	}
	}
\\
	& \leq 
	\av{
	f''(\phi_0)\nabla\phi_0\nabla {\rm E}^\phi_\epsilon
	}
		+\av{
			f'(\phi_0)\Delta {\rm E}^\phi_\epsilon
		}
\\&\qquad
	+\av{
		f'''(\phi_0)\av{\nabla\phi_0}^2{\rm E}^\phi_\epsilon
	}
	+\av{
		f''(\phi_0)\Delta\phi_0{\rm E}^\phi_\epsilon
	}
	\Bigr|
}{Repsphi0}
and hence
\bsplitl{
\int_{\Omega^\epsilon}
		\av{{\rm R}_\epsilon^\phi}
	\,d{\bf x}
	& \leq
		\N{f''(\cdot)}{L^\infty(I_\phi)}
		\N{\nabla\phi_0}{L^\infty(\Omega^\epsilon)}
		\N{\nabla {\rm E}^\phi_\epsilon}{L^2(\Omega^\epsilon)}
\\&\quad
		+
		\N{f'(\cdot)}{L^\infty(I_\phi)}
		\N{\Delta {\rm E}^\phi_\epsilon}{L^2(\Omega^\epsilon)}
\\&\quad
		+
		\N{f'''(\cdot)}{L^\infty(I_\phi)}
		\N{\nabla\phi_0}{L^\infty(I_\phi)}^2
		\N{{\rm E}^\phi_\epsilon}{L^2(\Omega^\epsilon}
\\&\quad
		+
		\N{f''(\cdot)}{L^\infty(I_\phi)}
		\N{\Delta\phi_0}{L^\infty(I_\phi)}
		\N{{\rm E}^\phi_\epsilon}{L^2(\Omega^\epsilon}
\\&
	\leq C\brkts{
				\epsilon^{1/2}
				+
				\N{{\rm E}^w_\epsilon}{L^2(\Omega^\epsilon}
	}
	\,,
}{Repsphi}
where 
$
\Ll{\Delta{\rm E}^\phi_\epsilon}
	\leq
	\Ll{{\rm E}^w_\epsilon}
	+\epsilon\Ll{{\rm F}^\phi_\epsilon}
$ thanks to \reff{ErEq}$_3$ 
}
and $I_\phi:=]\underline{\phi},\overline{\phi}[$ is the interval defined by the smallest 
real root $\underline{\phi}$ and $\overline{\phi}$ the largest real root of the polynomial free energy 
$F$ characterized by \reff{PolyDef}.

In order to control \cblu{the boundary contributions \reff{GwGphi}}, we apply a standard argument \cite{Bensoussans1978,Cioranescu1999} based on a cut-off function $\chi^\epsilon$ which is defined as follows,
\bsplitl{
\begin{cases}
\quad
\chi^\epsilon \in {\cg D}(\Omega^\epsilon)\,,&
\\\quad
\chi^\epsilon = 1
	& \textrm{if }{\rm dist}(x,\partial\Omega^\epsilon)\leq\epsilon\,,
\\\quad
\chi^\epsilon = 0
	& \textrm{if }{\rm dist}(x,\partial\Omega^\epsilon)\geq 2 \epsilon\,,
\\\quad
\N{\nabla \chi^\epsilon}{L^\infty(\Omega^\epsilon)}
	\leq \frac{C}{\epsilon}\,.
\end{cases}
}{CF}
\cblu{ 
We first look at  
${\rm G}_\epsilon^\phi\,.$
}
For $\eta^\phi_\epsilon :=  \chi^\epsilon {\rm G}^\phi_\epsilon$, we show that $\eta^\phi_\epsilon\in H^1(\Omega^\epsilon)$ and 
\bsplitl{
\N{\eta^\phi_\epsilon}{H^1(U^\epsilon)}
	\leq C\epsilon^{-1/2}\,,
}{G3Best}
where $U^\epsilon$ is the support of $\eta^\phi_\epsilon$ and forms a neighbourhood of $\partial\Omega^\epsilon$ of thickness $2\epsilon$. The regularity properties of $\xi^k_w$, 
$\xi_\phi^k$ and $\chi^\epsilon$ allow us to control $\eta_\epsilon^\phi$ as follows
\bsplitl{
\N{\eta^\phi_\epsilon}{H^1(U^\epsilon)}
	\leq C\brkts{
		\frac{1}{\epsilon}\N{\phi_0}{H^1(U^\epsilon)}
		+1
	}
	\,,
}{grdEta}
where $C$ is independent of $\epsilon$. Next, we use the result (\cite[Lemma 5.1, p.7]{Oleinik1992}), that is,
%
$
\N{\phi_0}{H^1(U^\epsilon)}
	\leq \epsilon^{1/2}C\N{\nabla\phi_0}{H^1(\Omega^\epsilon)}
	\,,
$
%
for a $C$ independent of $\epsilon$. Herewith, we established \reff{G3Best}. Using the 
trace theorem and the fact that $\eta^\phi_\epsilon=\cblu{{\rm G}^\phi_\epsilon}$ on $\partial\Omega^\epsilon$ allow 
us to obtain the estimate
\bsplitl{
\N{{\rm G}^\phi_\epsilon}{H^{1/2}(\partial\Omega^\epsilon)}
	= \N{\eta^\phi_\epsilon}{H^{1/2}(\partial\Omega^\epsilon)}
	\leq C\N{\eta^\phi_\epsilon}{H^1(\Omega^\epsilon)}
	= C\N{\eta^\phi_\epsilon}{H^1(U^\epsilon)}
	\,,
}{GepsCtrl}
which provides with \reff{G3Best} the bound
\bsplitl{
\N{{\rm G}_\epsilon^\phi}{H^{1/2}(\partial\Omega^\epsilon)}
	\leq C\epsilon^{-1/2}
	\,.
}{GepsCtrl1}
Applying the same arguments to ${\rm G}_\epsilon^w$ immediately leads to 
the corresponding bound
\bsplitl{
\N{{\rm G}_\epsilon^w}{H^{1/2}(\partial\Omega^\epsilon)}
	\leq C\epsilon^{-1/2}
	\,.
}{GWepsCtrl}

Next, we estimate \reff{ErEq}$_1$.
Testing \reff{ErEq}$_1$ with \cblu{$-\Delta {\rm E}_\epsilon^w={\cg A}_\epsilon{\rm E}_\epsilon^w $} provides 
\bsplitl{
\begin{cases}
\brkts{
		\partial_t \left(-\Delta\right)^{-1}{\rm E}_\epsilon^w, -\Delta {\rm E}_\epsilon^w
	}
	& = \brkts{
		\partial_t {\rm E}^w_\epsilon,{\rm E}_\epsilon^w
	}
	\cblu{
		+[BT1]
		-[BT2]
	}
	\,,
\\
-\lambda\brkts{
		{\cg B}_\epsilon {\rm E}_\epsilon^w,{\cg A}_\epsilon {\rm E}_\epsilon^w
	}
	& = -\lambda m\brkts{
		{\cg A}_\epsilon{\rm E}_\epsilon^w,
		{\cg A}_\epsilon{\rm E}_\epsilon^w
	}
	= -\lambda m\Ll{{\cg A}_\epsilon{\rm E}_\epsilon^w}^2
	\,,
\\
\brkts{
		{\cg B}_\epsilon {\rm E}^f_\epsilon,
		{\cg A}_\epsilon{\rm E}^w_\epsilon
	}
	& =
	m \Bigl(
		{\cg A}_\epsilon\Bigl\{
			f(\phi^\epsilon)-f(\phi_0)
			-\epsilon f'(\phi_0)\cblu{({\rm E}^\phi_\epsilon+\epsilon\phi_1)}
		\Bigr\},
		{\cg A}_\epsilon{\rm E}_\epsilon^w
	\Bigr)
	\,,
\\
\cblu{
\epsilon\brkts{ 
	{\rm R}^\phi_\epsilon, {\cal A}_\epsilon {\rm E}_\epsilon^w
	}
	}
	&\cblu{ 
	\leq  
		C(\kappa)\brkts{
			\epsilon
			+
			\N{{\rm E}_\epsilon^w}{L^2(\Omega^\epsilon)}^2
		}
	 	+\kappa/2\Ll{{\cal A}_\epsilon {\rm E}_\epsilon^w}^2
	}
\\
\epsilon\brkts{
	{\rm F}^w_\epsilon,{\cal A}_\epsilon {\rm E}_\epsilon^w
}
	& \leq 
		\epsilon C(\kappa)\Ll{{\rm F}_\epsilon^w}^2
		+\kappa/2\Ll{{\cal A}_\epsilon {\rm E}_\epsilon^w}^2
	\,,
\end{cases}
}{Eweps1}
where \cblu{we used \reff{Repsphi}} and the boundary terms $[BT1]$ and $[BT2]$ are bounded as 
follows
\bsplitl{
\cblu{
\av{[BT1]}
	}
	&\cblu{ :=
	\av{
	\int_{\partial\Omega^\epsilon}
		\partial_t(-\Delta)^{-1}\nabla_n {\rm E}^w_\epsilon {\rm E}^w_\epsilon
	\,d\sigma
	}
	=
	\av{ 
	\int_{\partial\Omega^\epsilon}
		\partial_t(-\Delta)^{-1}\epsilon {\rm G}^w_\epsilon {\rm E}^w_\epsilon
	\,d\sigma
	}
	}
\\&\qquad\cblu{
	\leq
		\epsilon C\N{\partial_t(-\Delta)^{-1}{\rm G}^w_\epsilon}{W^{l-3,2}(\Omega^\epsilon)}
		\N{{\rm E}^w_\epsilon}{H^1(\Omega^\epsilon)}
	\leq \epsilon^{1/2} C
	}
	\qquad\text{for }l\geq 4
	\,,
\\
\cblu{
\av{
[BT2]
	}
	}
	&\cblu{ := 
	\av{
	\int_{\partial\Omega^\epsilon}
		\partial_t(-\Delta)^{-1} {\rm E}^w_\epsilon \nabla_n {\rm E}^w_\epsilon		
	\,d\sigma
	}
	=
	\av{ 
	\int_{\partial\Omega^\epsilon}
		\partial_t(-\Delta)^{-1} {\rm E}^w_\epsilon \epsilon {\rm G}^w_\epsilon
	\,d\sigma
	}
	}
\cblu{
	\leq
		\epsilon^{1/2} C
	}
	\,.
}{BT}
\cblu{In \reff{BT}$_1$, 
we used Assumption C to assure in terms of regularity that 
$\phi_0=(-\Delta)^{-1}w_0\in C^1(0,T;W^{l,\infty}(\Omega^\epsilon))$ 
and hence the final bound is a consequence of the trace theorem 
and \reff{GWepsCtrl}. 
The same argument holds for \reff{BT}$_2$.} 
%

All this together then leads to the estimate
\bsplitl{
\frac{1}{2}\frac{d}{dt}\Ll{{\rm E}^w_\epsilon}^2
	& +(m\lambda-\kappa)\Ll{{\cg A}_\epsilon{\rm E}_\epsilon^w}^2
\\& 
	\leq C(m,\kappa)\Bigl(
		\av{\brkts{
			{\cg A}_\epsilon\ebrkts{
				f(\phi^\epsilon)
				-f(\phi_0)
			},{\cg A}_\epsilon{\rm E}_\epsilon^w
		}}
\\&\quad
	+\av{\brkts{
			{\cg A}_\epsilon[f'(\phi_0)\cblu{({\rm E}^\phi_\epsilon+\epsilon\phi_1)}],
			{\cg A}_\epsilon{\rm E}_\epsilon^w
		}}
	\Bigr)
\\&\quad
	+C(\kappa)\brkts{\epsilon+\N{{\rm E}^w_\epsilon}{L^2(\Omega^\epsilon)}^2}
	+\epsilon\Ll{{\rm F}_\epsilon^w}^2
	+\epsilon^{1/2} C
	\,,
}{Eweps2}
where the last summand reflects the boundary terms. 
In order to control the terms on the right-hand side in \reff{Eweps2}, we make use of 
the fact that $f(s)$ is a polynomial and satisfies \reff{Polyf}.
The first term in \reff{Eweps2} satisfies the following inequality
\bsplitl{
& 
\av{\brkts{
		{\cg A}_\epsilon\ebrkts{
			f(\phi^\epsilon)
			-f(\phi_0)
		},{\cg A}_\epsilon{\rm E}_\epsilon^w
	}}
\leq C\Bigl(
	\av{\brkts{
		\brcs{
			f''(\phi^\epsilon)
			-f''(\phi_0)
		}\av{\nabla\phi^\epsilon}^2,
		{\cg A}_\epsilon{\rm E}_\epsilon^w
	}}
\\&
	+\av{\brkts{
		f''(\phi_0)\nabla\brkts{
			\phi^\epsilon
			+\phi_0
		}\nabla\brkts{
			\phi^\epsilon
			-\phi_0
		},
		{\cg A}_\epsilon{\rm E}_\epsilon^w
	}}
\\&
	+\av{\brkts{
		\brcs{
			f'(\phi^\epsilon)
			-f'(\phi_0)
		}\Delta\phi^\epsilon,
		{\cg A}_\epsilon{\rm E}_\epsilon^w
	}}
	+\av{\brkts{
		f'(\phi_0)\Delta(\phi^\epsilon-\phi_0),
		{\cg A}_\epsilon{\rm E}_\epsilon^w
	}}
	\Bigr)
\,.
}{FrEnDi}
Before we proceed, we estimate the terms on the right-hand side in \reff{FrEnDi}:

\medskip

\emph{1st term in \reff{FrEnDi}:} We first note that with the remainder term in Taylor series 
we obtain 
\bsplitl{
&\av{f''(\phi^\epsilon)
		-f''(\phi_0)
	}
	\leq 
		\sup_{\theta\in I_\phi}f'''(\theta)\av{
			\phi^\epsilon-\phi_0
	}
	\leq 
		\N{f'''(\cdot)}{L^\infty(I_\phi)}
		\av{\phi^\epsilon-\phi_0}
\\&\qquad
	\leq 
		\N{f'''(\cdot)}{L^\infty(I_\phi)}
		\brkts{
			\av{{\rm E}_\epsilon^\phi}
			+\epsilon\av{\phi_1}
		}
\\&\qquad
	\leq
		\N{f'''(\cdot)}{L^\infty(I_\phi)}
		\brkts{
			\av{{\rm E}_\epsilon^\phi}
			+\epsilon\av{(\pmb{\xi}_\phi\cdot\nabla_x)\phi_0}
	}
\\&\qquad
	\leq
		\N{f'''(\cdot)}{L^\infty(I_\phi)}
		\brkts{
			\av{{\rm E}_\epsilon^\phi}
			+\epsilon C
	}
	\,,
}{1FrEnDi}
where $I_\phi:=]\underline{\phi},\overline{\phi}[$ is the interval defined by the smallest 
real root $\underline{\phi}$ and $\overline{\phi}$ the largest real root of the polynomial free energy 
$F$ characterized by \reff{PolyDef}. We used Assumption C in \reff{1FrEnDi}.
Herewith, we can estimate the first term (e.g. in $d=3$) as follows
\bsplitl{
&
\av{\brkts{
		\brcs{
			f''(\phi^\epsilon)
			-f''(\phi_0)
		}\av{\nabla\phi^\epsilon}^2,
		{\cg A}_\epsilon{\rm E}_\epsilon^w
	}}
\\&
	\leq C\N{f'''(\cdot)}{L^\infty(I_\phi)}
		(\N{{\rm E}_\epsilon^\phi}{L^6}+\epsilon)
		\N{\nabla\phi^\epsilon}{L^6}^2
		\Ll{{\cg A}_\epsilon {\rm E}_\epsilon^w}
\\&
	\leq C\N{f'''(\cdot)}{L^\infty(I_\phi)}
		(\N{{\rm E}_\epsilon^\phi}{H^1}+\epsilon)
		\N{\nabla\phi^\epsilon}{H^1}^2
		\Ll{{\cg A}_\epsilon {\rm E}_\epsilon^w}
\\&
	\leq  C(\kappa)(\N{{\rm E}_\epsilon^\phi}{H^1}^2+\epsilon^2)
		+\kappa\Ll{{\cg A}_\epsilon {\rm E}_\epsilon^w}^2
\\&
	\leq C(\kappa) \brkts{
			\Ll{{\rm E}_\epsilon^w}^2
			+ \epsilon^2 (2+\epsilon^{1/2})^2
		}
		+\kappa\Ll{{\cg A}_\epsilon {\rm E}_\epsilon^w}^2
	\,.
}{1FrEnDi1}
\\
\emph{2nd term in \reff{FrEnDi}:} With Sobolev inequalities, e.g. 
Fridrichs' inequality in the perforated domain case \cite{Chechkin2007}, and the identity 
$\phi^\epsilon-\phi_0={\rm E}_\epsilon^\phi+\epsilon\phi_1$ we 
obtain the following estimate
\bsplitl{
&\av{\brkts{
		f''(\phi_0)\nabla\brkts{
			\phi^\epsilon
			+\phi_0
		}\nabla\brkts{
			\phi^\epsilon
			-\phi_0
		},
		{\cg A}_\epsilon{\rm E}_\epsilon^w
	}}
	\leq
		C(T)
		\Bigl(
			\N{\nabla\phi^\epsilon}{L^6}
\\&\qquad\quad
			+\N{\nabla\phi_0}{L^6}
		\Bigr)\Bigl(
			\N{\nabla{\rm E}_\epsilon^\phi}{L^3}
			+\epsilon\N{\nabla\phi_1}{L^3}
		\Bigr)
		\Ll{{\cg A}_\epsilon {\rm E}_\epsilon^w}
\\&\qquad
	\leq C(T,\kappa) \Bigl(
			\Ll{{\rm E}_\epsilon^w}^2
			+ \epsilon^2
		\Bigr)
		+\kappa\Ll{{\cg A}_\epsilon {\rm E}_\epsilon^w}^2
	\,,
}{2FrEnDi}
where we again used Assumption C and classical regularity results 
from the elliptic PDE theory.
\\
\emph{3rd term in \reff{FrEnDi}:} Following the same ideas as for the \emph{1st term} estimated 
in \reff{1FrEnDi1}, we immediately get the bound
\bsplitl{
&\av{\brkts{
		\brcs{
			f'(\phi^\epsilon)
			-f'(\phi_0)
		}\Delta\phi^\epsilon,
		{\cg A}_\epsilon{\rm E}_\epsilon^w
	}}
\\&\quad
	\leq C(\Omega,T,\kappa) \brkts{
			\Ll{{\rm E}_\epsilon^w}^2
			+ \epsilon^2 
		}
		+\kappa\Ll{{\cg A}_\epsilon {\rm E}_\epsilon^w}^2
	\,.
}{3FrEnDi}
\\
\emph{4th term in \reff{FrEnDi}:} The last term can finally be controlled as follows
\bsplitl{
&\av{\brkts{
		f'(\phi_0)\Delta(\phi^\epsilon-\phi_0),
		{\cg A}_\epsilon{\rm E}_\epsilon^w
	}}
\\&\quad
	\leq
	C(\Omega,T)
	\Bigl(
		\Ll{\Delta{\rm E}_\epsilon^\phi}
		+\epsilon\Ll{\Delta\phi_1}
	\Bigr)\Ll{{\cg A}_\epsilon{\rm E}_\epsilon^w}
\\&\quad
	\leq 
	C(\Omega,T,\kappa)
	\Bigl(
		\brkts{
			\Ll{{\rm E}_\epsilon^w}^2
			+\epsilon^2\Ll{{\rm F}_\epsilon^\phi}^2
		}
		+\epsilon^2
	\Bigr)+\kappa\Ll{{\cg A}_\epsilon{\rm E}_\epsilon^w}^2
	\,,
}{4FrEnDi}
where we again used Assumption C and classical regularity results 
from the theory of elliptic PDEs.

\medskip

Back to controlling \reff{Eweps2}, \cblu{it leaves to control 
the second term on the right-hand side, i.e., 
$
\av{\brkts{{\cal A}_\epsilon[f'(\phi_0)({\rm E}^\phi_\epsilon+\epsilon\phi_1)],{\cal A}_\epsilon E_\epsilon^w}}
$. 
We have 
\bsplitl{
\av{\brkts{{\cal A}_\epsilon[f'(\phi_0)({\rm E}^\phi_\epsilon+\epsilon\phi_1)],{\cal A}_\epsilon E_\epsilon^w}}
	& \leq
	C\brkts{
		\epsilon
		+\N{{\rm E}^\phi_\epsilon}{L^2(\Omega^\epsilon)}
		+\N{\nabla{\rm E}^\phi_\epsilon}{L^2(\Omega^\epsilon)}
		+\N{{\rm E}^w_\epsilon}{L^2(\Omega^\epsilon)}
	}
\\&
	\leq 
	C\brkts{
		\epsilon 
		+\N{{\rm E}^w_\epsilon}{L^2(\Omega^\epsilon)}	
	}
	\,,
}{ErEs2ndTB}
where we used the facts that 
\bsplitl{
{\cal A}_\epsilon[f'(\phi_0)({\rm E}^\phi_\epsilon+\epsilon\phi_1)]
	& =-\Delta[f'(\phi_0)({\rm E}^\phi_\epsilon+\epsilon\phi_1)]
\\&
	=-f'''(\phi_0)|\nabla\phi_0|^2({\rm E}^\phi_\epsilon+\epsilon\phi_1)
	-f''(\phi_0)\Delta\phi_0({\rm E}^\phi_\epsilon+\epsilon\phi_1)
\\&
	-f''(\phi_0)\nabla\phi_0\nabla({\rm E}^\phi_\epsilon+\epsilon\phi_1)
	-f'(\phi_0)\Delta({\rm E}^\phi_\epsilon+\epsilon\phi_1)
	\,,
}{2ndTRw}
Assumption C, and error equation \reff{ErEq}$_3$.
}

\medskip

\cblu{Hence, with the previously derived bounds 
\reff{FrEnDi} and \reff{ErEs2ndTB} we obtain,}
\bsplitl{
\frac{d}{dt}\Ll{E_\epsilon^w}^2
	&
	+2(m\lambda-\kappa)\Ll{{\cal A}_\epsilon E_\epsilon^w}^2
\\
	&
	\leq C(m,\kappa,\lambda,\Omega,T)\brkts{
		\Ll{E_\epsilon^w}^2 
		+ A(\epsilon)
	}
	+8\kappa\Ll{{\cal A}_\epsilon E^w_\epsilon}^2
	\,,
}{ctrlOnEw}
where 
$
A(\epsilon):= \brkts{
		\epsilon^2(2+\epsilon^{1/2})^2+\epsilon^2+\epsilon+\epsilon^{1/2}
	}
\,.$ 
A consideration of 
$\frac{d}{dt}\brkts{{\rm exp}\brkts{-Ct}\Ll{{\rm E}_\epsilon^w}}$ 
leads after some rewriting to the following bound, 
\bsplitl{
\Ll{{\rm E}_\epsilon^w(\cdot,T)}^2
	&
	\leq 
	{\rm exp}\brkts{CT}C
	A(\epsilon)
	\,,
}{Eweps3}
which quarantees the control of \reff{Eweps2}. Herewith, we are 
also in the position to derive a bound on 
$\Ll{{\cal A}_\epsilon^w E_\epsilon^w}^2$ 
after integrating \reff{ctrlOnEw} over time, that means, 
\bsplitl{
\Ll{E^w_\epsilon}^2(t)
	& 
	+2\brkts{m\lambda-n\kappa}\int_0^t\Ll{{\cal A}_\epsilon E^w_\epsilon}^2(s)\,ds
\\&
	\leq C\brkts{
		{\rm exp}\brkts{Ct}
	}A(\epsilon)t
	\,,
}{lapEw}
for $n\in\mathbb{N}$ finite. 
\hfill $\square$

%% file: conclusion8.tex
Based on a microscopic porous media formulation \reff{PeMoPr}, we derived
upscaled/ homogenized phase-field equations for general free energy
densities. We gave a rigorous justification of this new effective macroscopic
equations for a class of polynomial free energies which include the widely
used double-well potential. The porous materials considered here can be
represented by a periodic covering of a single reference cell $Y$ which
accounts for the pore geometry. It is well-known that transport as well as
fluid flow in porous media lead to high-dimensional computational problems,
since the mesh size needs to be much smaller than the heterogeneity
$\epsilon:=\frac{\ell}{L}$, the ratio of the characteristic length scale of
the pores $\ell$ over the size of macroscopic porous medium $L$.
We rigorously derived qualitative error estimates for the approximation error between the solution
of the effective macroscopic problem \reff{pmWrThm} and the solution of the
fully resolved microscopic equation \reff{PeMoPr}. We also recovered the
classical error behavior from homogenization of elliptic
problems based on a truncation not resolving boundary effects  
in the context of fourth order phase field problems.

This error quantification is of fundamental interest in applications as it
provides guidance on the applicability of the new effective macroscopic
phase-field formulation in dependence of the heterogeneity $\epsilon>0$
defined by the heterogeneous material under consideration. Our dimensionally
reduced phase field formulation can also be seen as the precursor to an
effective and systematic computational strategy where microscopic properties
such as the geometry and wall characteristics (e.g. wetting properties) of a
reference pore enter the macroscopic description in an effective manner, thus
avoiding a full numerical resolution of the finer details of the porous
structure.

There are a number of interesting mathematical-physical questions related to
the analysis presented here. For instance: (i) The error estimates still
require rather high regularity assumptions (Assumption C) for which 
Lemma \ref{lem:ApEs} seems to provide currently the best available 
estimates; (ii)
Moreover, the error behavior in time in \reff{ThmErEs} seems not optimal;
(iii) And a more physically motivated question is: ``What is the influence of
the pore or material geometry on phase transformations in heterogeneous media
such as composites and porous materials and does the effective formulation 
capture such geometry dependent phenomena?''

Finally, we believe that, due to the popularity and the wide range of
applicability of phase-field equations, the new effective macroscopic
formulation, could ultimately serve as a promising computational tool in
material-chemical- physical sciences and engineering. In particular, the
effective phase-field formulation \reff{pmWrThm} could form the basis for a
promising new direction for modelling multiphase flow in porous media without
making use of Darcy's law.

We shall examine these and related questions in future studies.